\documentclass[aps,prb,fleqn,amssymb]{revtex4}
\usepackage{hyperref}
\usepackage{graphicx}
\usepackage{amsmath}
\usepackage{bm}

\newcommand{\beq}{\begin{equation}}
\newcommand{\eeq}{\end{equation}}
\newcommand{\bD}{{\bf D}}
\newcommand{\bI}{{\bf I}}
\newcommand{\bQ}{{\bf Q}}

\newcommand{\bU}{{\bf U}}
\newcommand{\bSigma}{{\bf \Sigma}}
\newcommand{\bff}{{\bf f}}
\newcommand{\bk}{{\bf k}}
\newcommand{\br}{{\bf r}}
\newcommand{\bu}{{\bf u}}
\newcommand{\bx}{{\bf x}}
\newcommand{\bp}{{\bf p}}
\newcommand{\bn}{{\bf n}}
\newcommand{\bq}{{\bf q}}
\newcommand{\by}{{\bf y}}

\begin{document}
\title{Multiscale modeling and simulation of microtubule/motor protein
  assemblies}

\author{Tong Gao$^{1}$, Robert Blackwell$^{2}$, Matthew
  A. Glaser$^{2}$, M. D. Betterton$^{2}$, Michael J. Shelley$^{1}$}
\affiliation{$^1$Courant Institute of Mathematical Sciences, New York
  University, New York, NY 10012 \\
  $^2$Department of Physics, University of Colorado, Boulder, CO
  80309}

\begin{abstract}\noindent
  Microtubules and motor proteins self organize into biologically
  important assemblies including the mitotic spindle and the
  centrosomal microtubule array. Outside of cells, microtubule-motor
  mixtures can form novel active liquid-crystalline materials driven
  out of equilibrium by ATP-consuming motor proteins.  Microscopic
  motor activity causes polarity-dependent interactions between motor
  proteins and microtubules, but how these interactions yield such
  larger-scale dynamical behavior such as complex flows and defect
  dynamics is not well understood. We develop a multiscale theory for
  microtubule-motor systems in which Brownian dynamics simulations of
  polar microtubules driven by motors are used to study microscopic
  organization and stresses created by motor-mediated microtubule
  interactions. We identify polarity-sorting and crosslink tether
  relaxation as two polar-specific sources of active destabilizing
  stress. We then develop a continuum Doi-Onsager model that captures
  polarity sorting and the hydrodynamic flows generated by these
  polar-specific active stresses. In simulations of active nematic
  flows on immersed surfaces, the active stresses drive turbulent flow
  dynamics and continuous generation and annihilation of disclination
  defects.  The dynamics follow from two instabilities, and accounting
  for the immersed nature of the experiment yields unambiguous
  characteristic length and time scales. When turning off the
  hydrodynamics in the Doi-Onsager model, we capture formation of
  polar lanes as observed in the Brownian dynamics simulation.
\end{abstract}

\date{\today}
\maketitle

\section{Introduction}
Active matter, the novel class of nonequilibrium materials
made up of self-driven constituents, presents scientific challenges to
our understanding of material properties and has the potential to
provide new technologies such as autonomously moving and
self-healing materials. Examples of active matter include flocks of
birds\cite{cavagna10}, swarms of swimming bacteria\cite{zhang10} or
self-propelled colloidal particles\cite{bricard13}, and the cellular
cytoskeleton and cytoskeletal
extracts\cite{nedelec97,surrey01,schaller10,sanchez12}. Despite their
differences in composition and length scale, these diverse systems
show common features absent in equilibrium systems, including
collective motion, nonequilibrium ordering transitions, and anomalous
fluctuations and mechanical properties\cite{ramaswamy10,marchetti13}.
Understanding and predicting the properties of active matter require
new theoretical approaches and models applicable to
far-from-equilibrium, internally driven systems.

Mixtures of cytoskeletal filaments and motors are an important class
of active matter that can be reconstituted outside the cell to form
novel materials.  Filaments driven into self-organized patterns such
as vortices and asters are reminiscent of structures observed in
cells\cite{nedelec97,surrey01,tanaka-takiguchi04,janson07,bendix08,pinot09,hentrich10,thoresen11,gordon12,schaller10}.
In earlier experiments, filaments were driven into static
self-organized patterns such as vortices and asters, reminiscent of
structures observed {\it in vivo}. In recent experiments, active
networks were formed of microtubules (MTs) and synthetic multimeric
kinesin-1 motor complexes, with the aid of a
depletant\cite{sanchez12,henkin14,keber14}. In the presence of ATP,
motor complexes can bind pairs of MTs and walk along MTs towards their
plus-ends.  When suspended in bulk, depletion interactions drove the
formation of extended, highly ordered MT bundles characterized by
bundle extension and fracture, and correlated with spontaneous
large-scale fluid flows\cite{sanchez12,keber14}. When MT bundles were
adsorbed onto an oil-water interface, they formed a dense, nematically
ordered surface state and exhibited an active nematic phase
characterized by the spontaneous generation and annihilation of
disclination defect pairs\cite{sanchez12}.

Theoretical studies
\cite{nakazawa96,kruse00,lee01,kruse03,liverpool03,sanakararaman04,aranson05,Liverpool05,ziebert05,ahmadi06,giomi08,zemel09}
have investigated aspects of these active-matter systems at different
scales, from the dynamics and mechanical properties of filament
bundles to macroscopic behavior and stability of active suspensions.
Inspired by the experiments of Sanchez {\it et al.} \cite{sanchez12},
both Giomi {\it et al.} \cite{giomi13,giomi14} and Thampi {\it et al.}
\cite{TGY2013,thampi14a,thampi14b,thampi14c} have studied liquid crystal
hydrodynamic models with fluid flow driven by an apolar active stress
\cite{simha02,Voituriez05}. In these rather general models the precise
origins of the active stress driving the system are
unidentified. Giomi {\it et al.}  developed a theory for the speed at
which defects move apart in active nematics, assuming the presence of
a defect pair as an initial condition. Thampi {\it et al.}  found an
activity-independent velocity-velocity correlation length, as found in
the bulk flow measurements of Sanchez {\it et al.}, and studied defect
dynamics in 2D simulations. These models reproduce qualitative
features of the experiments.  However, MT/motor-protein interactions
are intrinsically polar, and how these polarity-dependent microscopic
interactions manifest themselves at meso- or macroscopic scales is
still unknown.  Thus one theoretical challenge is how to resolve
microscopic interactions between constituents in order to predict
macroscopic material properties. While particle-based simulations can
represent microscopic interactions in detail, computational cost
typically limits cross-scale studies. Continuum models are more
tractable for describing dynamics at large scales, but can be
difficult to connect to the microscopic dynamics quantitatively.

Here we construct a multi-scale model that identifies the sources of
destabilizing active stresses, and study their consequences in a
large-scale model \cite{GBGBS2015}. We first perform detailed, hybrid
Brownian dynamics-kinetic Monte Carlo (BD-kMC) simulations which
incorporate excluded-volume interactions between model MTs, thermal
fluctuations, explicit motors with binding/unbinding kinetics that
satisfy detailed balance, and a force-velocity relation. Active
extensile stress is generated from polarity sorting of anti-aligned
MTs, and from crosslink relaxation of polar-aligned MTs.  It also
provides coefficients for polar-specific active stresses for a
kinetic theory that incorporates polarity sorting and long-range
hydrodynamic interactions, using a similar approach as that used to
describe bacterial suspensions
\cite{saintillan08,saintillan08a,Wolgemuth08,subramanian09,Wensink12,ESS2013},
where hydrodynamic instabilities lead to large-scale collective
motions including jets and vortices
\cite{pedley92,koch11,simha02,saintillan08,saintillan08a,sokolov09,kurtuldu11}.
We use this model to study actively streaming nematic states on an
immersed surface, as in the Sanchez {\it et al.}
experiments\cite{sanchez12}.  Numerical experiments demonstrate
dynamics strikingly similar to the experiments, with large-scale
turbulent-like fluid flows and the persistent production and
annihilation of defects. We correlate the defect dynamics with
specific flow structures and with active stresses. We identify the
hydrodynamic instability of nearly 1D coherent ``cracks'' as being
source of the persistent dynamics. When turning off the induced
background surface flow in the kinetic model, we capture the formation
of polar lanes observed in the BD-kMC simulation.

\section{The Microscopic Model}
\label{micromodeling}

\begin{figure*}
  \begin{center}
     \includegraphics[scale=0.8]{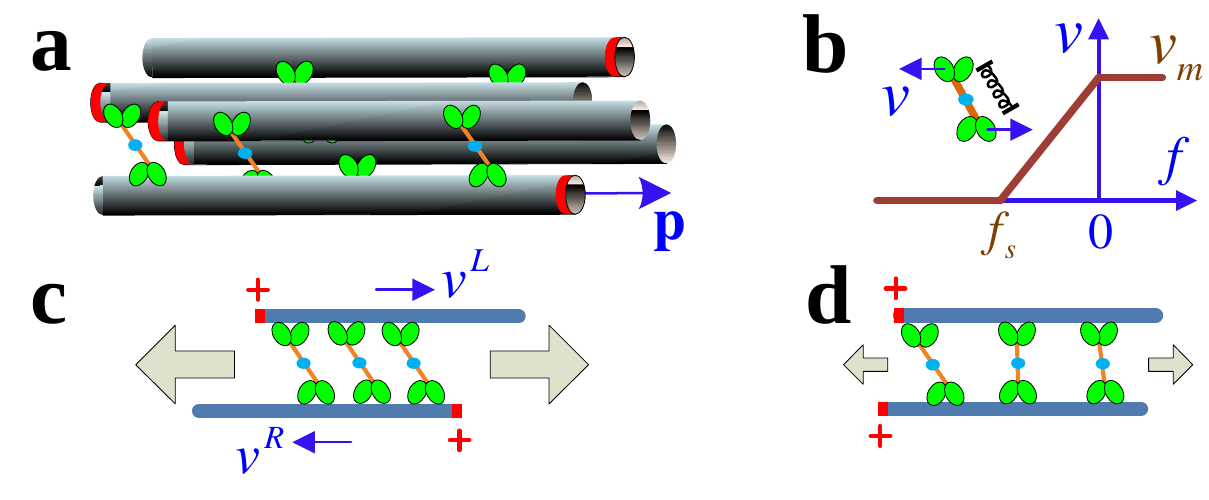}
  \end{center}
  \caption{(a) Schematic of a cluster of polar-aligned and
    anti-aligned MTs, with plus ends marked by red rings. Motors walk
    on neighboring MTs, and (b) exert spring-like forces with a
    piecewise linear force-velocity relation. (c) An anti-aligned MT
    pair. (d) A polar-aligned MT pair.  Grey arrows characterize the
    magnitude of the extensile stress.}
  \label{schematic}
\end{figure*}

Figure~\ref{schematic} outlines the basic physical picture that
underlies both our BD-kMC simulations and the continuum kinetic model.
Consider an immersed suspension of polar MTs, each with a plus-end
oriented director $\bp$, and all of the same length $l$ and diameter
$b$ (Fig.~\ref{schematic}a). Adjacent MTs are coupled by plus-end
directed crosslinking motors consisting of one motor head on each MT
connected by a tether that responds as a spring to stretching
(Fig.~\ref{schematic}b). The motor on each crosslink endpoint moves
with a linear force-velocity relation \cite{visscher99}:
$v=v_m\max(0,\min(1,1+f/f_s))$, where $f$ is the magnitude of the
crosslinking force, $v_m$ is the maximum translocation velocity, and
$f_s$ is the stall force.  For a nematically aligned suspension there
are two basic types of MT pair interaction. For polar anti-aligned MTs
(Fig.~\ref{schematic}c) the motors on each end of an active crosslink
move in opposite directions, stretching the tether. This creates
forces on each MT that, acting against fluid drag, slide the MTs
relative to each other towards their minus-ends. This process is
termed {\it polarity sorting} \cite{nakazawa96}. Conversely, for
polar-aligned MTs the motors on each end of the crosslink move in the
same direction, there is little or no net sliding, and the tether
pulling on the leading motor causes stretched tethers to relax
(Fig.~\ref{schematic}d).

\section{Brownian dynamics-kinetic Monte Carlo model and simulations}
We first perform 2D Brownian dynamics-kinetic Monte Carlo (BD-kMC)
simulations of MTs driven by explicit motors with binding/unbinding
kinetics. The main purpose is to quantify local MT pair interactions,
with long-ranged hydrodynamics neglected due to its high computational
cost.  Our model is related to previous simulations of filaments with
crosslinking motors \cite{nedelec07,head11,head14}, but new in our
work are algorithmic improvements for handling crosslinks and neglect
of filament elasticity that allow us to more accurately treat the
statistical mechanics of crosslinking motors, simulate larger systems
and measure the stress tensor.

The particle-based BD-kMC simulations use a simple, tractable model of
active biomolecular assemblies that capture key physical features,
including excluded volume interactions between filaments, attractive
and sliding forces exerted by motors, and the thermodynamics and
kinetics of crosslinking motor binding and unbinding. Filaments (MTs)
are represented as perfectly rigid rods (discorectangles in 2D) of
length $l$ and diameter $b$ that undergo Brownian dynamics. Forces and
torques on the filaments occur due to motor-mediated forces,
particle-particle repulsion, friction, and thermal forces.  To
simulate the Brownian motion of filaments, we adopt the computational
scheme of Tao {\it et al.} \cite{tao05}, which has been used
successfully in simulations of concentrated solutions of
high-aspect-ratio particles. In this scheme, the filament centers of
mass equations of motion are
\begin{equation}
\label{eq:browniancm}
{\bf x}_i(t + \delta t) = {\bf x}_i(t) + {\bf \Gamma}_i^{-1}(t) \cdot
{\bf F}_i(t) \delta t + \delta {\bf x}_i(t)
\end{equation}
for all filaments $i$, where the random displacement
$\delta {\bf x}_i(t)$ is Gaussian-distributed and anisotropic, with
variance
\begin{equation}
  \left\langle \delta {\bf x}_i(t) \delta {\bf x}_i(t) \right\rangle =
  2 k_B T {\bf \Gamma}_i^{-1}(t) \delta t.
\end{equation}
In the above, $k_B$ is
Boltzmann's constant and $T$ is the absolute temperature.
Here ${\bf \Gamma}_i^{-1}(t)$ is the inverse friction tensor
\begin{equation}
{\bf \Gamma}_i^{-1}(t) = \gamma_\parallel^{-1} {\bf p}_i(t) {\bf
  p}_i(t) + \gamma_\perp^{-1} \left[ {\bf I} - {\bf p}_i(t) {\bf
    p}_i(t) \right],
\end{equation}
where $\gamma_\parallel$ and $\gamma_\perp$ are the parallel and
perpendicular drag coefficients of the rod, and ${\bf F}_i(t)$ is the
systematic (deterministic) force on particle $i$. The equations of
motion for particle reorientation are
 \begin{equation}
\label{eq:brownianrot}
{\bf p}_i(t + \delta t) = {\bf p}_i(t) + {1 \over {\gamma_r}} {\bf
  T}_i(t) \times {\bf p}_i(t) \delta t + \delta {\bf p}_i(t),
\end{equation}
where $\gamma_r$ is the rotational drag coefficient, ${\bf T}_i(t)$ is
the systematic torque on particle $i$, and the random reorientation
$\delta {\bf p}_i(t)$ is Gaussian-distributed, with variance
\begin{equation}
\left\langle \delta {\bf p}_i(t) \delta {\bf p}_i(t) \right\rangle = 2
k_B T {1 \over {\gamma_r}} \left[ {\bf I} - {\bf p}_i(t) {\bf p}_i(t)
\right] \delta t.
\end{equation}

The Weeks-Chandler-Andersen (WCA) potential between rods is
\begin{widetext}
\begin{equation}
u_{\rm wca}(r_{\rm min}) = \left\{
\begin{array}{ll}
4 \epsilon \left[ \left( {b \over r_{\rm min}} \right)^{12} -
  \left( {b \over r_{\rm min}} \right)^6 \right] + \epsilon, &
r_{\rm min} < 2^{1/6} b \\
0, & r_{\rm min} \geq 2^{1/6} b,
\end{array}
\right.
\end{equation}
\end{widetext}
where $\epsilon = k_B T$, $r_{\rm min}$ is the minimum distance between the two finite
line segments that define the filament axis, and $\epsilon$ sets the
energy scale of the potential. Note that $r_{\rm min}$ is an implicit
function of the center-of-mass positions and orientations of the two
interacting MTs.  For this value of $\epsilon$, the typical distance of closest approach between
rods is comparable to $b$, and the thermodynamic properties closely
resemble those of hard rods with aspect ratio $l/b$, a model that is
well-characterized both in 2D \cite{bates00} and 3D
\cite{bolhuis97,mcgrother96}.

Because the Brownian dynamics scheme involves random particle
displacements and rotations, close contacts between rods that produce
large forces and torques occasionally occur, leading to instabilities
in the dynamics. Such instabilities are avoided by softening the WCA
potential at short distances to keep the resulting forces and torques
within reasonable bounds \cite{tao05}. At the same time, we adjust the
integration timestep to ensure that pairs of interacting particles
probe the softened region of the potential infrequently, so that
excluded volume effects are properly accounted for.

The frictional forces are orientation dependent: translational
diffusion is characterized by two diffusion constants, $D_\perp$ and
$D_\parallel$, which describe diffusion perpendicular and parallel to
the rod axis, respectively, and $D_r$ is the rotational diffusion
coefficient. For spherocylinders where $a=l/b+1$, the diffusion
coefficients are \cite{lowen94}
\begin{equation}
  D_\parallel = {{k_B T} \over {2 \pi \eta (l+1)}} \left ( \\
  \ln a - 0.207 + 0.980 / a - 0.133 a^2 \right),
\end{equation}
\begin{equation}
  D_\perp = {{k_B T} \over {4 \pi \eta (l+1)}} \left ( \\
  \ln a + 0.839 + 0.185 / a + 0.233 / a^2 \right ),
\end{equation}
and
\begin{equation}
  D_r = {{3 k_B T} \over {\pi \eta (l+1)^3}} \left ( \\
  \ln a - 0.662 + 0.917 / a - 0.050 / a^2 \right ).
\end{equation}
Here $\eta$ is the fluid viscosity and $k_B T$ is the temperature in
energy units.  Note that $D_\parallel$ is approximately a factor of
two larger than $D_\perp$.

To model motor-mediated interactions and activity, we implement a
semi-grand canonical ensemble in which a reservoir of motors is
maintained in diffusive contact at a fixed chemical potential
$\mu_{m}$ with filaments to/from which they can bind/unbind. The
motors are assumed to be noninteracting both in solution and in the
bound state, so the motor reservoir can be treated as an ideal
solution, and there is no steric interference among bound motors.
Bound motors have a free energy $u_m(r_m)$, where $r_m$ is the
extension of the motor tether, which depends implicitly on the
relative positions and orientations of the two filaments to which the
motors is attached and on the positions of the points of attachment of
the motor on the filament axes. We treat motor attachment (detachment)
as a one-step process in which motors bind to (unbind from) two
filaments simultaneously, and we assume a binding rate of
\begin{equation}
k_{\rm on}(r) = k_0 e^{- \beta u_m(r)}
\label{eq:binding_rate}:
\end{equation}
and an unbinding rate of
\begin{equation}
k_{\rm off}(r) = k_0,
\label{eq:unbinding_rate}
\end{equation}
where $\beta = (k_B T)^{-1}$ is the inverse temperature in energy
units. This choice of binding and unbinding rates ensures that the
correct equilibrium distribution is recovered for static
(non-translocating) crosslinks, is a convenient choice from a
computational standpoint, and has been used
previously\cite{lenz14}. Given a distribution of motors bound to
filaments, we compute the forces and torques exerted on MTs by
differentiating $u_m(r_m)$ with respect to the filament positions and
orientations.  As discussed in section \ref{micromodeling}, the
endpoints of bound motors translocate toward the plus ends of the MTs
to which they are attached with a force-dependent velocity. Motors
unbind immediately upon reaching the plus end of either of the two
filaments to which they are attached.

Because the motor unbinding rate is $k_0$, independent of motor tether
extension, the probability that a given motor unbinds in a time
interval $\delta t$ is $p = k_0 \delta t$, and the average number of
motors that unbind in $\delta t$ is
$\langle N_{\rm d} \rangle = k_0 \delta t N_m$, where
$N_m$ is the current number of bound motors. The number
$N_{\rm d}$ of motors that unbind in a time interval $\delta t$
follows a binomial distribution,
\begin{equation}
{\cal P}(N_{\rm d}) = {N_m \choose N_{\rm d}} p^{N_{\rm d}}
(1-p)^{N_m - N_{\rm d}}.
\label{eq:binomial}
\end{equation}
In one timestep we remove $N_{\rm d}$ randomly selected motors, where
$N_{\rm d}$ is determined by sampling from the binomial distribution.

The kinetic MC procedure for motor binding is involved, because the
rate of motor binding depends on motor tether extension, which in turn
depends on the relative positions and orientations of the two MTs to
which the motor is attached and on the positions of the points of
attachment of the motor along the filament axes. To compute the
relative probability and rate of motor binding to specific binding
sites on a given pair of filaments we consider the statistical
mechanics of the filament/motor system in the equilibrium limit of
non-translocating crosslinks.  The semi-grand canonical partition
function of the filament/motor system is
\begin{equation}
  \Xi(N,V,T,\mu_m) = \sum_{N_m = 0}^\infty z_{\rm
    c}^{N_m} Z(N,V,T,N_m),
\end{equation}
where $z_m = e^{\beta \mu_m}$ is the fugacity of the
motor reservoir and $N_m$ is the number of bound motors.
Here, $Z(N,V,T,N_m)$ is the canonical partition function of a
system of $N$ filaments and $N_m$ bound motors,
\begin{widetext}
\begin{equation}
Z(N,V,T,N_m) = {{1} \over {N!}} \int d{\bf x}^N d{\bf p}^N e^{-
  \beta U({\bf x}^N, {\bf p}^N)}
\left[ {{1} \over {N_m!}} \left( q_1 \right)^{N_m} \right]
\label{eq:canonical}
\end{equation}
\end{widetext}
where $({\bf x}^N, {\bf p}^N) = ({\bf x}_1, {\bf x}_2,...{\bf x}_N,
{\bf p}_1, {\bf p}_2, ..., {\bf p}_N)$ labels the particle positions
and orientations, $U({\bf x}^N, {\bf p}^N)$ is the filament potential
energy, including interparticle interactions and external potentials,
$q_1$ is the single-motor partition function, and interactions
between bound motors have been neglected. The single-motor
partition function depends on the filament positions ${\bf x}^N$and
orientations ${\bf p}^N$, i.e., $q_1 = q_1({\bf x}^N, {\bf p}^N)$.
Substituting Eq.~(\ref{eq:canonical}) into the grand partition function
and carrying out the summation over $N_m$ leads to
\begin{widetext}
\begin{equation}
\Xi(N,V,T,\mu_m) = {{1} \over {N!}} \int d{\bf x}^N d{\bf p}^N
e^{- \beta [ U({\bf x}^N, {\bf p}^N) + U_m({\bf x}^N, {\bf
    p}^N)]}
\label{eq:grand_canonical}
\end{equation}
\end{widetext}
where
\begin{equation}
  U_m({\bf x}^N, {\bf p}^N) = - {{z_m} \over {\beta}}
  q_1({\bf x}^N, {\bf p}^N).
\end{equation}
In the limit in which the rate of motor binding and unbinding is
large compared with the filament diffusion rate (adiabatic limit),
$U_m$ plays the role of an effective motor-mediated filament
interaction potential that depends on the chemical potential of the
reservoir. Static-crosslink-mediated interactions are generally
attractive and short-ranged, and bear a strong resemblance to
depletion-type potentials \cite{bolhuis97}.

The single-motor partition function $q_1$ can be written as a sum
of pairwise partition functions,
\begin{equation}
q_1({\bf x}^N, {\bf p}^N) = \sum_{i < j}^N q_{ij} ({\bf x}_i, {\bf
  p}_i, {\bf x}_j, {\bf p}_j)
\end{equation}
where the sum ranges over all distinct pairs of filaments, and the
pairwise partition function $q_{ij}$ is
\begin{widetext}
\begin{equation}
  q_{ij} ({\bf x}_i, {\bf p}_i, {\bf x}_j, {\bf p}_j) = \rho^2
  \int_{-l/2}^{l/2} d s_i \int_{-l/2}^{l/2} ds_j\ e^{-\beta
    u_m[r_m(s_i,s_j; {\bf x}_i, {\bf p}_i, {\bf x}_j, {\bf
      u}_j)]}
\label{eq:pair_partition_function}
\end{equation}
\end{widetext}
Here the integration variables $s_i$ and $s_j$ parametrize the
positions of motor endpoints on filaments $i$ and $j$, respectively,
$r_m$ is length of a motor between points specified by $s_i$ and
$s_j$, and $\rho$ is the linear density of binding sites on a single
filament. Then we can write the effective motor potential as the sum
of pairwise effective interactions,
\begin{equation}
U_m({\bf x}^N, {\bf p}^N) = \sum_{i < j}^N U_{ij} ({\bf x}_i,
{\bf p}_i, {\bf x}_j, {\bf p}_j)
\label{eq:effective_potential}
\end{equation}
where
\begin{equation}
U_{ij} ({\bf x}_i, {\bf p}_i, {\bf x}_j, {\bf p}_j) = - {{z_m}
  \over {\beta}}  q_{ij} ({\bf x}_i, {\bf p}_i, {\bf x}_j, {\bf p}_j)
\label{eq:cross_pair_poten}
\end{equation}
is the effective motor-mediated pair potential in the adiabatic limit.
Insertion of motors with the correct relative statistical weight in a
kinetic MC procedure requires evaluation of the pairwise partition
function $q_{ij}$ (Eq.~(\ref{eq:pair_partition_function})) for all
pairs of filaments.  If the motor energy $u_m$ increases rapidly
(e.g., quadratically) with increasing motor extension, the partition
function $q_{ij}$ (and the corresponding adiabatic effective potential
$U_{ij}$) falls off rapidly with increasing minimum distance between
filament axes, and is non-negligible only for pairs of filaments in
close proximity. Thus, the pairwise partition function is analogous to
a short-range interaction potential, and the usual techniques for
efficient handling of short-range interactions (e.g., neighbor lists)
can be applied. To efficiently evaluate the double integral in
Eq.~(\ref{eq:pair_partition_function}), note that for motors modeled
as zero-equilibrium-length harmonic springs, the integrand can be
expressed as a sum of bivariate normal distributions. Then $q_{ij}$
reduces to a sum of cumulative bivariate normal distributions, which
can be rapidly evaluated using standard numerical procedures
\cite{vesely06}.

To proceed further, we consider the statistical mechanics of the
motor subsystem for {\em fixed} filament positions and
orientations. The grand partition function for the motor subsystem
is given by
\begin{equation}
\Xi_m(N,T,\mu_m) = \sum_{N_m = 0}^\infty {{z_{\rm
      m}^{N_m}} \over {N_m!}} \left( q_1 \right)^{N_{\rm
    m}},
\label{eq:grand_canonical_motor}
\end{equation}
and the equilibrium number of bound motors for a given filament
configuration is
\begin{equation}
\left\langle N_m \right\rangle = \beta^{-1} {{\partial} \over
  {\partial \mu_m}} \ln \Xi_m
 =  \sum_{i < j}^N \left\langle N_{ij} \right\rangle,
\end{equation}
where $\langle N_{ij} \rangle$ is the average number of motors
between filaments $i$ and $j$,
\begin{widetext}
\begin{equation}
  \left\langle N_{ij} ({\bf x}_i, {\bf p}_i, {\bf x}_j, {\bf p}_j)
  \right\rangle = z_m \rho^2 \int_{-l/2}^{l/2} ds_i
  \int_{-l/2}^{l/2} ds_j \ e^{-\beta u_{m}[r_m(s_i,s_j; {\bf
      r}_i, {\bf p}_i, {\bf x}_j, {\bf p}_j)]}.
\label{eq:avg_pair_motors}
\end{equation}
\end{widetext}
Note that $\langle N_{ij} \rangle = z_m q_{ij}$, so the problem of
computing $\langle N_{ij} \rangle$ is equivalent to that of computing
$q_{ij}$. Introducing the explicit form of quadratic potential for
harmonic motors $u_m(r_m) = -u_0 + \frac{1}{2} K r_m^2$ leads to
\begin{equation}
\left\langle N_{ij} \right\rangle = z_m \rho^2 e^{\beta u_0}
\int_{-l/2}^{l/2} ds_i \int_{-l/2}^{l/2} ds_j\ e^{- \alpha
  r_m^2(s_i,s_j)},
\label{eq:avg_pair_motors_2}
\end{equation}
where $\alpha = \beta K / 2$, and where the implicit dependence of $r_m$
on filament coordinates has been suppressed.

The average number of motors that bind to filaments in a time interval
$\delta t$ is
\begin{equation}
\left\langle N_{\rm a} \right\rangle = k_0 \delta t \left\langle
  N_m \right\rangle = k_0 \delta t \sum_{i < j}^N \left\langle
  N_{ij} \right\rangle.
\end{equation}
As above, the number $N_{\rm a}$ of motors that bind in the interval
$\delta t$ follows a Poisson distribution,
\begin{equation}
{\cal P}(N_{\rm a}) = {{\left\langle N_{\rm a} \right\rangle^{N_{\rm a}}
    e^{-N_{\rm a}}} \over {N_{\rm a}!}}.
\label{eq:poisson}
\end{equation}
In the kinetic MC cycle, the number of bound motors
$N_{\rm a}$ is drawn from this distribution, and $N_{\rm a}$
motors are inserted by first selecting pairs of filaments with
relative probability ${\cal P}_{ij} = \langle N_{ij} \rangle / \langle
N_m \rangle$ and then sampling from the appropriate bivariate
normal distribution to choose motor endpoints that lie on the
selected pair of filaments.

The overall hybrid BD-kMC procedure thus consists of the following
steps:
\begin{enumerate}
\item Compute forces and torques on MTs, and evolve MT positions and
  orientations $\delta t$ forward in time according to the Brownian
  dynamics equations of motion (Eqs.~(\ref{eq:browniancm}) and
  (\ref{eq:brownianrot})).
\item Displace each motor endpoint by $v \delta t$ along the MT to
  which it is attached with translocation velocity $v$ given by the
  force-velocity relation.
\item Determine the number $N_{\rm d}$ of motors that unbind in the
  time interval $\delta t$ by drawing from a binomial distribution
  (Eq.~(\ref{eq:binomial})), and remove this number of motors at
  random.
\item Compute average number of bound motors $\langle N_{ij} \rangle$
  for all pairs of MTs (Eq.~(\ref{eq:avg_pair_motors})) and determine
  the number $N_{\rm a}$ of motors that bind in the time interval
  $\delta t$ by drawing from a Poisson distribution
  (Eq.~(\ref{eq:poisson})). Randomly select $N_{\rm a}$ pairs of MTs
  with relative probability
  $\langle N_{ij} \rangle / \sum_{i < j}^N \left\langle N_{ij}
  \right\rangle$,
  and insert a motor between each selected pair of MTs by sampling
  from a bivariate normal distribution.
\end{enumerate}

The properties of the model depend on seven dimensionless parameters
(tables \ref{params} and \ref{params-dimensionless}): the MT aspect
ratio $r = l/b$, the MT packing fraction $\phi$, the range of motor
mediated interaction $R_m = [k_B T / (K b^2)]^{1/2}$, the motor
concentration $c = z_m \rho^2 b^2 e^{u_0/(k_B T)}$, the motor run
length $\ell= v/(k_0 l)$, the motor stall force
$f = f_{\rm s} b / (k_B T)$, and the Peclet number (the ratio of
translocation and diffusion rates) $Pe = v \eta b / (k_B T)$.  With
current methods, it becomes more computationally expensive to simulate
systems with MTs of high aspect ratio (e.g., $r>10$). The computation
time scales approximately as $r^3$. If $r$ doubles, the linear
dimension of the the box in the longitudinal direction must be doubled
to study the same number of rods. We use square boxes to avoid any
loss of information upon nematic director reorientation. Therefore the
number of rods scales as $r^2$.  Longer rods also have slower
dynamics, because the translational and rotational mobilities go as
$1/r$ to leading order.  Therefore the timescale to reach steady state
scales approximately linearly in $r$.  We present here results of
simulations with $r=10$ for which we performed simulations of
relatively large systems for long times over a wide range of
parameters. A more limited investigation of longer rods reveal
qualitatively similar behavior.

\begin{table}
\begin{center}
  \begin{tabular}{|c|p{5cm}|p{3.5cm}|p{5cm}|}
    \hline
    \multicolumn{4}{|c|}{Parameter values}\\
    \hline
    Quantity & Parameter & Value  & Notes  \\
    \hline
    $k_B T$  & Thermal energy & 4.11 $\times 10^{-21}$ J & Room
                                                           temperature \\
    \hline
    $l$    & MT length & 250 nm & Chosen  \\
    \hline
    $b$    & MT diameter & 25 nm & Ref \onlinecite{alberts07} \\
    \hline
    $\epsilon$  & Energy scale of the WCA potential   & $k_B T$ &  Refs \onlinecite{bates00,bolhuis97,mcgrother96}   \\
    \hline
    $\eta$  & Fluid viscosity   & 1.0 Pa s &  Cytoplasmic viscosity, ref
                                             \onlinecite{Wirtz09}   \\
    \hline
    $\rho$  & Linear density of motor  binding sites along MT & -- &
                                                                     Appears only in dimensionless concentration  \\
    \hline
    $\mu_m$  & Motor chemical potential & -- &  Appears only in
                                               dimensionless concentration   \\
    \hline
    $u_0$  & Motor binding free energy &--  &  Appears only in
                                              dimensionless concentration    \\
    \hline
    $v_w$ & Motor speed (zero force) & Reference 4.5 $\mu$m/s, range
                                       0.14--18 $\mu$m/s & Of order 1
                                                           $\mu$m/s,
                                                           ref
                                                           \onlinecite{visscher99}\\
    \hline
    $k_0$  & Unbinding rate of motors & 28.1 s$^{-1}$&  Processivity
                                                       of 160 nm, ref \onlinecite{schnitzer00}\\
    \hline
    $f_s$  & Stall force  &  1 pN &  Ref \onlinecite{visscher99} \\
    \hline
    $K$  & Motor spring constant  & 0.013 pN/nm & Decreased from ref
                                                  \onlinecite{Coppin95}
                                                  to give appropriate range of motor-mediated interaction for
                                                  zero-equilibrium-length
                                                  springs\\
    \hline
  \end{tabular}
  \caption{Parameter values of the BD-kMC simulation.}
  \label{params}
\end{center}
\end{table}

\begin{table}
\begin{center}

  \begin{tabular}{|c|l|p{2.5 cm}|p{5cm}|}
    \hline
    \multicolumn{4}{|c|}{Dimensionless parameter values}\\
    \hline
    Quantity & Parameter & Value & Notes  \\
    \hline
    $\phi$  & MT packing fraction  & 0.54 &  Chosen to give nematic
    state at equilibrium in the absence of motors   \\
    \hline
    $r = l/b$& MT aspect ratio & 10 & \\ \hline
    $c=\rho^2 b^2 e^{\beta (\mu_m+u_0)}$  & Motor
    concentration  & 1 &  Chosen to give average of 2 motors per
    nearby MT pair   \\  \hline
    $R_m = \sqrt{k_B T / (K
      b^2)}$ & Range of motor interaction& $1/\sqrt{2}$&
    Chosen to be of order 1 for a short-range interaction\\ \hline
    $\ell= v_w/(k_0 l)$ & Motor run length & Reference 0.64, range
    0.2--12.8 & Motor-induced active
    stresses are largest when $\ell$ is of order 1. \\ \hline
    $f = f_{\rm s} b / (k_B T)$ & Motor stall force & 6 &\\ \hline
    $Pe = v_w \eta  b / (k_B T)$ & Peclet number & Reference 0.68,
    range 0.02--2.7&\\
    \hline
  \end{tabular}
  \caption{Dimensionless groups of the BD-kMC simulation.}
  \label{params-dimensionless}
\end{center}
\end{table}

\subsection{Measurement}
The dynamics and stresses experienced by individual MTs depends
strongly on their local environment, in particular on the relative
polarity of neighboring MTs. To identify sub-populations of MTs with
distinct local environments, we define a local polar orientational
order parameter 
\begin{equation}
m_i = \frac{\sum_{j \neq i}^N {\bf p}_i \cdot {\bf p}_j \ q_{ij}
}{\sum_{j \neq i}^N q_{ij}},
\label{eq:local_polar_order_parameter}
\end{equation}
where $q_{ij}$ is the motor pair partition function defined above.
Since $q_{ij}$ falls off rapidly with increasing pair separation, only
near neighbors of particle $i$ are included in the sums in
Eq.~(\ref{eq:local_polar_order_parameter}).  The polar order parameter
$m_i$ ranges from $-1$ (MT $i$ surrounded by neighbors of opposite
polarity) to $1$ (MT $i$ surrounded by neighbors of the same
polarity).

The osmotic stress tensor of a periodic system of $N$ interacting
MTs at temperature $T$ in a $d$-dimensional volume $V$ is given
by
\begin{equation}
{\bf \Sigma} = {{N k_B T} \over {V}} {\bf I} + {{1} \over {V}} \left\langle
  {\bf W} \right\rangle,
\label{eq:pressure_tensor}
\end{equation}
where the first and second terms on the right-hand side represent the
ideal gas and interaction contributions, respectively, ${\bf I}$ is
the unit tensor, and ${\bf W}$ is the virial tensor,
\begin{equation}
{\bf W} = \sum_{i < j}^N {\bf r}_{ij} {\bf F}_{ij}.
\end{equation}
where the sum ranges over all interacting pairs of MTs. The
angular brackets in Eqn.~\eqref{eq:pressure_tensor} denote an average
over time. Here we have assumed that the temperature of the system is
isotropic and well-defined, so that
\begin{equation}
\left\langle \sum_{i=1}^N {{{\bf P}_i {\bf P}_i} \over {m_{\rm MT}}}
\right\rangle = {{N k_B T} \over {V}} {\bf I},
\end{equation}
where ${\bf P}_i$ is the momentum of MT $i$ and $m_{\rm MT}$ is the MT
mass (here assumed the same for all MTs). Filaments have momentum
based on their instantaneous movements on short time-scales. This
motion is in thermal equilibrium with the background fluid, connecting
molecular motion to Brownian motion. While this relation is clearly
true in the equilibrium case, it's less obvious that this it holds for
active MT/motor systems.  However, a purely mechanical definition of
osmotic pressure leads to the same expression even for nonequilibrium
particle suspensions in the low Reynolds number hydrodynamic regime
\cite{brady93}, and we will assume that Eq.~(\ref{eq:pressure_tensor})
holds in the following discussion.

The isotropic pressure is defined as
\begin{equation}
\left\langle \Pi \right\rangle = {1 \over d} \sum_{j = 1}^d
\left\langle \Sigma_{j j} \right\rangle,
\end{equation}
The average extensile stress is
\begin{equation} {\bf{\Sigma}}_b = {1 \over {d - 1}} \sum_{j = 1}^{d -
    1} \left\langle \Sigma_{d d} \right\rangle - \left\langle
    \Sigma_{j j} \right\rangle,
\end{equation}
where the $d$ direction corresponds to the average nematic director
orientation. We further resolve the stress tensor into contributions
from sub-populations of MTs, for example according to the local polar
order parameter $m_i$ introduced above. This can be done by writing
the total virial as the sum of contributions from individual MTs,
\begin{equation}
{\bf W} = \sum_{i = 1}^N {\bf W}_i,
\end{equation}
where
\begin{equation}
{\bf W}_i = {1 \over 2} \sum_{j \neq i}^N {\bf r}_{ij} {\bf F}_{ij}.
\label{eq:single_particle_virial}
\end{equation}

To calculate the pair extensile stresslet as a function of the local
polar order $m_i$, we calculate the virial per spherocylinder. At a
given time point, each interaction gives an associated virial
contribution for the pair. The single-MT virial contribution is taken
to be half of the pair's contribution. Contributions from forces for
all interacting partners are summed up to give the virial contribution
for each MT. Similarly, the local polar order parameter $m_i$ is
calculated for each MT. Then the virial anisotropy contribution per MT
in the nematic reference frame is determined based on its local polar
order. After repeating for all time points, the histogram is
normalized, leading to the calculation of the extensile pair stresslet
per MT as a function of $m_i$.

To calculate the extensile pair stresslet in bulk simulations, we
consider interacting MTs only. At each time point, the total number of
interactions is calculated by summing the number of pairs for which
there is a nonzero force. The total parallel and antiparallel virials
in the director reference frame are calculated.  Any interactions
between pairs with ${\bf p}_i \cdot {\bf p}_j>0$ contribute to the
polar-aligned virial, and the remainder contribute to the anti-aligned
virial. This measurements is time averaged and the extensile pair
stresslet calculated by dividing the average virial anisotropy by the
average number of interactions.

\subsection{Extensile stress and its origins}

\begin{figure*}
   \includegraphics[width=164 mm]{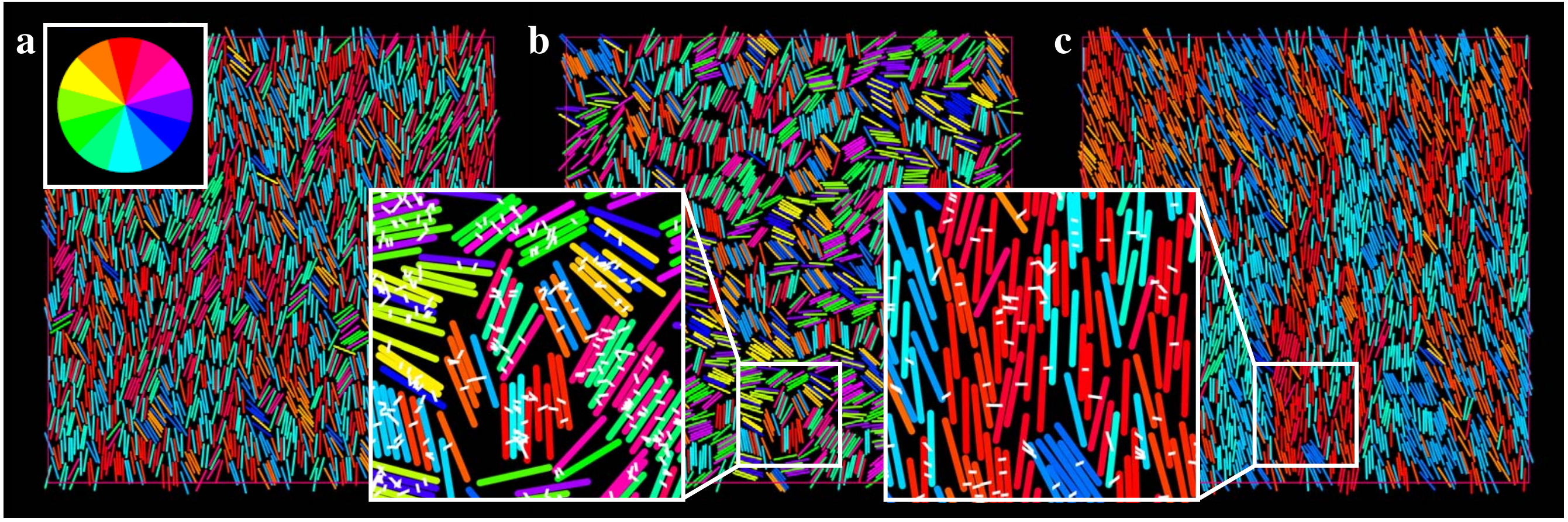}
   \caption{Snapshots of the BD-kMC particle
     simulations. Insets are zoomed views with motors explicitly shown in
     white. (a) System with no motors, illustrating the 2D nematic
     state. (b) An equilibrium system with static crosslinkers
     exhibits MT bundling due to short-range crosslink-induced
     attraction. (c) An active system with motors exhibits active
     flows and formation of polar lanes. }
 \label{figure02}
\end{figure*}

\begin{figure*}
   \includegraphics[width=164 mm]{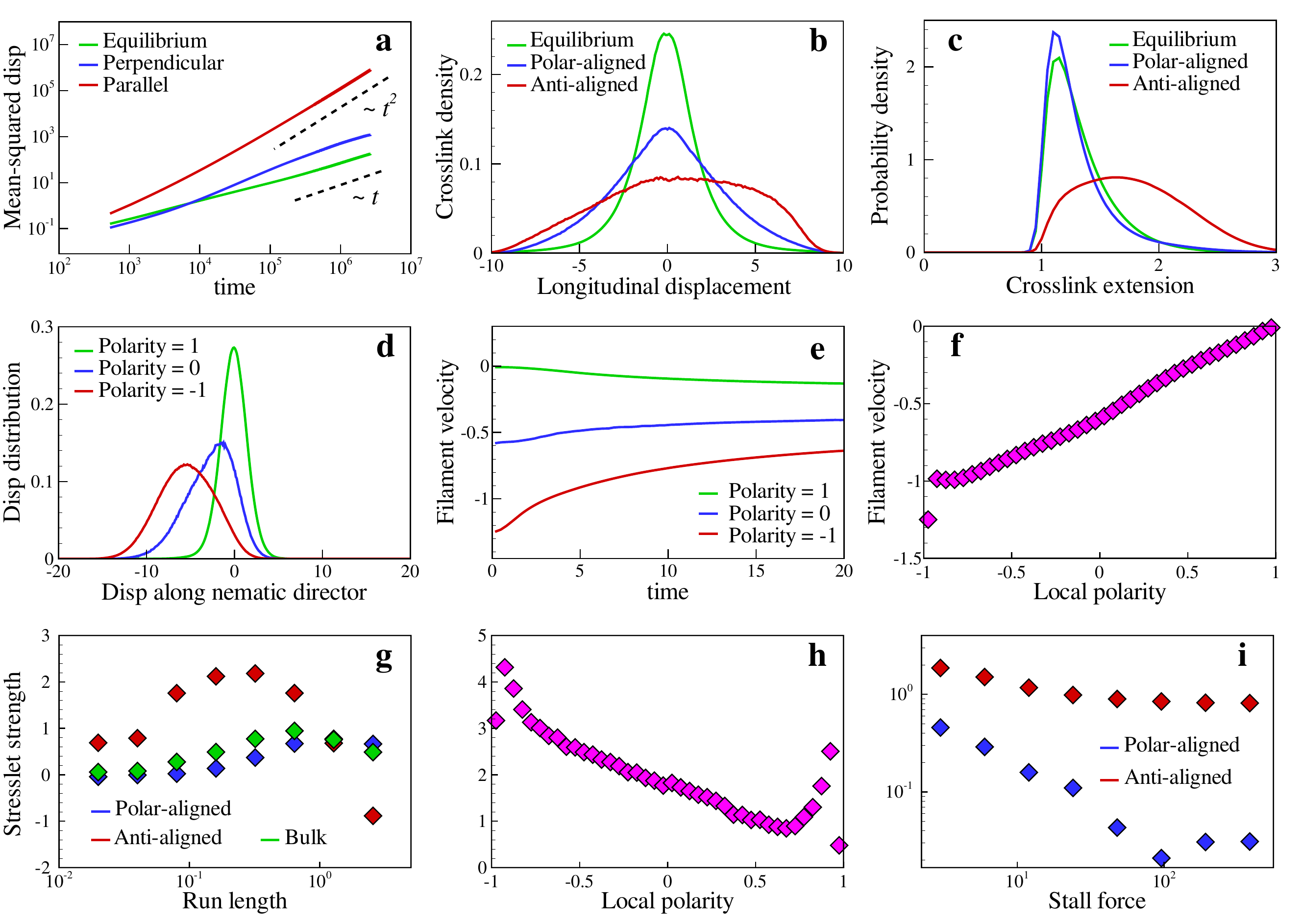}
   \caption{Measurements of BD-kMC simulations. (a) Mean-squared
     deviation of MTs as a function of time. (b) Mean velocity of
     filaments along nematic director as a function of time $t$ for
     different initial polar environments $m_i$. (c) Histogram of
     motor extension $r_m$, broken into contributions from
     polar-aligned and anti-aligned pairs in the active case. (d)
     Histogram of motor occupancy as a function of the particle
     filament longitudinal displacement $s_{ij}$, broken into
     contributions from polar-aligned and anti-aligned pairs in the
     active case.  (e) Histogram of filament displacements at time
     separation $t = 4.98$ for various initial polar environments
     $m_i$. (f) Variation of average instantaneous velocity of
     filaments along the nematic director in time for different
     initial polar environments $m_i$. (g) Variation of extensile pair
     stresslet $S$ with motor run length $\ell$, showing results from
     the entire bulk simulation and contributions of polar-aligned and
     polar-antialigned pairs. (h) Variation of extensile pair
     stresslet with local polar environment $m_i$. (i) Variation of
     extensile pair stresslet with motor stall force from simulations
     of isolated, perfectly parallel filament pairs. }
 \label{figure03}
\end{figure*}

Figure \ref{figure02} illustrates the long-time behavior of MT
suspensions in the BD-kMC simulation model (also see video S1). Fig.~\ref{figure02}a shows
a simulation of MTs interacting only through thermal fluctuations and
steric interactions (without motors). The system develops a 2D nematic
state consistent with previous work\cite{bates00}.
Figure~\ref{figure02}b shows the result of adding immobile
crosslinkers with full binding/unbinding kinetics. The system shows MT
bundling due to short-range crosslink-induced attraction.
Figure~\ref{figure02}c shows the behavior with motors. The system now
shows active MT flows driven by polarity sorting, leading to the
formation of polar lanes (domains of MTs with similar polarity). These
polar lanes are highly dynamic and show large fluctuations. The
mean-squared displacement of MT position as a function of time shows
diffusive behavior at long-times in the equilibrium cases
(Figs.~\ref{figure02}a and b) and for active MTs when measured
perpendicular to the average alignment direction. For motion parallel
to the average alignment direction, the active MT mean-squared
displacement is superdiffusive and nearly ballistic at long times
(Fig.~\ref{figure03}a).

We characterized the dynamical properties of bound motors for
polar-aligned and anti-aligned MT pairs. For two MTs labeled $i$ and
$j$ with orientations $\bp_i$ and $\bp_j$ and center-of-mass
diplacement $\br_{ij}$, we define the pair's longitudinal displacement
by
$s_{ij}={1 \over
  2}\br_{ij}\cdot[\bp_i+\operatorname{sgn}(\bp_i\cdot\bp_j) \bp_j ]$.
For anti-aligned MT pairs ($\bp_i\cdot\bp_j<0$) undergoing
motor-driven relative sliding, $s_{ij}$ is negative when the MT pair
is contracting (minus-ends closer than plus-ends), and becomes
positive when the MT pair is extending (plus-ends closer than
minus-ends; see Fig.~\ref{schematic}). When crosslinks are immobile or
for motors on polar-aligned MTs ($\bp_i\cdot\bp_j\geq 0$), the
distribution of motors as a function of $s_{ij}$ is symmetric
(Fig.~\ref{figure03}a).  However for motors on anti-aligned MTs, the
distribution of motors skews toward positive values of $s_{ij}$: more
motors are bound during the extensile motion of the pair. This
asymmetry occurs because of the translocation of the motors toward the
MT plus-ends. This biases MT pairs toward extension, yielding an
extensile stress that drives active flows (see below).

The distribution of motor extension alters significantly when
crosslinks translocate (Fig.~\ref{figure03}c). The minimum value of
$r_m$ is approximately 1 due to excluded-volume interactions between
MTs. For polar-aligned pairs, the distribution is shifted toward
\textit{smaller} extensions than in the equilibrium case due to
nonequilibrium tether relaxation, with important implications for the
generation of extensile stress, as discussed below. For anti-aligned
pairs, the distribution is shifted toward positive extension due to
oppositely directed motor motion; this combination of motor extension
and motion applies active forces that drive polarity sorting.

We measured the displacement distributions and average velocities of
MTs along the nematic director and found that both are strongly
correlated with an MT's initial local polar environment. Defining the
nematic director $\hat{\bf n}$, we calculated MT displacement
distributions in time along the projection of the local filament
orientation vector onto the nematic vector:
$y(t) = \textrm{sgn}(\hat{\bf n} \cdot {\bf p}_i(t_0))~\hat{\bf
  n}\cdot [{\bf r}_i(t+t_0) - {\bf r}_i(t)]$.
In order to examine dynamical behavior on timescales comparable to the
diffusion timescale, we grouped the MT displacements at a lag time of
$t=4.98$ (chosen to clearly illustrate the different distributions)
and their initial polar environment $m_i(t_0) \approx (-1,0,1)$. For
MTs in an initially polar environment ($m_i(t_0) \approx 1)$, the
displacement distribution is approximately Gaussian with mean near
zero, consistent with diffusive-like dynamics
(Fig.~\ref{figure03}d). For MTs in an initially anti-polar environment
($m_i(t_0) \approx -1$), we again find an approximately Gaussian
displacement distribution, but the mean is shifted toward the MT's
minus end (Fig.~\ref{figure03}d). This profile is consistent with
drift plus diffusion dynamics. For more mixed initial environments
($m_i(t_0) \approx 0$), we find that the dynamics are more complicated
and are not likely described by a simple drift and diffusion model
(Fig.~\ref{figure03}d). MTs in initially mixed or anti-polar
environments exhibit significant displacements toward their minus ends
due to anti-polar sliding.

To further examine the polarity-dependent MT movements, we measured
instantaneous MT velocity component along the nematic director,
$dy/dt$ at $t=0$. Velocities of MTs are not constant because MTs
experience relatively rapid changes in the polarity of their
neightbors. MTs in initially anti-polar environments tend to slow down
rapidly, indicating that they move into more mixed environments, while
MTs in polar or mixed environments tend to maintain their velocities
for longer times. Filaments in polar environments have velocities near
zero. (Fig.~\ref{figure03}e). The instantaneous velocity depends
approximately linearly on the local polar environment, as expected
when filament movements are determined mainly by polarity sorting
(Fig.~\ref{figure03}f).

We measured the time-averaged bulk stress tensor $\bSigma_b$ for our
active particle system, and find that, over a wide range of
parameters, $\bSigma_b$ is anisotropic with larger components in the
average MT alignment direction than in the perpendicular
direction. That is, since the MT alignment direction is essentially
${\hat\by}$, the stress difference $\Sigma_b^{yy}-\Sigma_b^{xx}$ is
positive, which corresponds to an extensile stress.  Static
crosslinkers or no motors (Fig.~\ref{figure02}a, b) yield an isotropic
$\bSigma_b$.  The stress difference can be expressed as the sum of
pair interactions between nearby MTs, with each $ij$ pair contributing
a stresslet $S_{ij}$ (with units of force$\times$length), prior to
division by the bulk volume. We have characterized how the stresslet
varies with system parameters and configurations. The average pair
stresslet $S$ increases with the motor speed $v_m$ up to a maximum
where the typical motor run length is the MT length
(Fig.~\ref{figure03}g). Increasing $v_m$ further leads to decreasing
$S$ because the motors rapidly move to the ends of the MTs and
unbind. To understand the origins of extensile stress, we studied how
$S$ varies with the local polar environment $m_i$
\eqref{eq:local_polar_order_parameter}. The stresslet is largest when
$m_i$ is near $-1$, suggesting that polarity sorting is the dominant
source of pairwise extensile stress (Fig.~\ref{figure03}h).  As $m_i$
increases, $S$ drops with approximate linearity, at least away from
the two isolated peaks that close examination show originate
through strong steric interactions of nearly parallel MTs. Nearly, but
not exactly, parallel MTs experience aligning torques due to
motor-mediated attraction; the resulting steric collisions tend to
promote pair extension that increases the extensile stress for
nearly-aligned pairs (relative to perfectly aligned pairs).

The extensile stress from anti-aligned pair interactions arises from
asymmetries during polarity sorting: if an MT pair begins sliding when
the two minus-ends touch ($s_{ij}=-10$) and slides under a force
proportional to pair overlap until the two plus-ends meet
($s_{ij}=10$), then the contractile motion would perfectly balance the
extensile motion and the total extensile stress would be
zero\cite{kruse00,Liverpool05,lenz12,lenz14}. In our simulations we
observe two effects that break this symmetry. First, MTs are unlikely
to begin interacting exactly when their minus ends are touching,
decreasing the range of negative $s_{ij}$ over which sliding occurs.
Second, more motors are bound on average during extensional motion (so
that $s_{ij}>0$; see Fig.~\ref{figure03}b).

We also find the surprising and counterintuitive result that $S$
remains positive even when $m_i$ is near 1, that is, for polar-aligned
pairs of MTs. This effect occurs due to an interplay between motor
motion and excluded-volume interactions. We propose that the effect
can be understood by considering equilibrium and nonequilibrium motor
relaxation. For immobile motors, the system is at equilibrium and the
stress tensor is isotropic; attractive interactions due to motors are
balanced by excluded volume interactions and thermal fluctuations, and
the system is at mechanical equilibrium. When motors are active,
stress anisotropy becomes possible due to the nonequilibrium nature of
the motor force-velocity relation. The tether of a longitudinally
stretched motor pulls back on the leading motor, slowing it, and pulls
forward on the trailing motor. Hence, the motor relaxes its
longitudinal extension. This effect is observable in
Fig.~\ref{figure03}c as a slight but significant shift in the
distribution of motor extension toward smaller values relative to the
equilibrium case. As a result, the motor-induced contractile stress
along the MT alignment direction is decreased, while there is no
change in the transverse stress induced by motors. This leads to a net
anisotropic extensile stress in the alignment direction.  In this
scenario, we would predict that if the motors had a force-independent
velocity, the polar-aligned extensile stress would vanish because the
longitudinal motor extension would be unable to relax. We tested this
prediction by studying how $S$ varies as stall force increases for
simulations of perfectly aligned (unable to rotate) isolated filament
pairs. We find that the extensile stress changes little with stall
force for anti-aligned MT pairs. However, for polar-aligned MT pairs
the extensile stress drops as stall force increases and goes to zero
for infinite stall force (which corresponds to force-independent
velocity, Fig.~\ref{figure03}i). When effects of filament rotation are
also included, the results are more subtle; we find that the interplay
of filament rotation and motor activity can induce extensile stress
for polar-aligned pairs in bulk simulations even for infinite stall
force.

While the extensile stress due to polar-aligned MT pairs is typically
a factor of 2--5 smaller than for anti-aligned pairs, when measured
per pair (Fig.~\ref{figure03}h), polarity sorting and the tendency to
form polar lanes (Fig.~\ref{figure02}c) lead to larger numbers of
polar-aligned MT pairs than of anti-aligned. In our BD-kMC
simulations, which lack the effect of hydrodynamics, the overall
contributions of polar-aligned and anti-aligned pairs to the extensile
stress are comparable.

\section{Continuum kinetic theory}
The BD-kMC simulations show how polar-specific MT-pair interactions
give rise to extensile active stresses. To study the effect of
hydrodynamic interactions and to make analytical predictions we have
developed a Doi-Onsager theory \cite{doi88} similar to those used to
describe the dynamics of motile rod suspensions
\cite{saintillan08,saintillan08a,ESS2013}. The theory's fluxes and
active stresses arise from polar-aligned and anti-aligned MT pair
interactions produced by active motors. These stresses induce chaotic
flows driven by the formation of disclination defects.

\subsection{Dynamics of polarity sorting}
\begin{figure*}
  \begin{center}
     \includegraphics[scale=1.1]{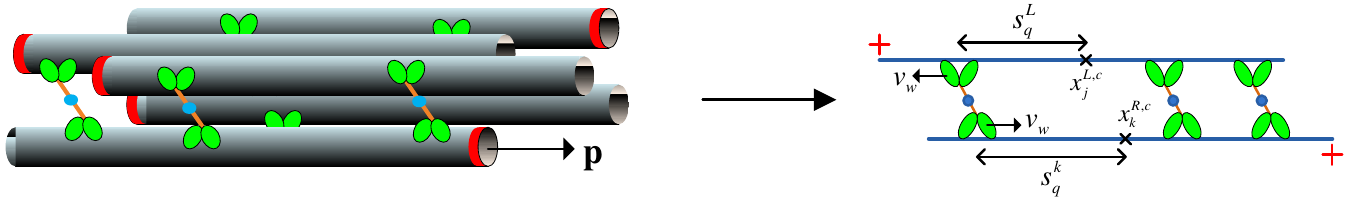}
  \end{center}
  \caption{Schematic for a cluster of MTs undergoing polarity
    sorting. The plus-ends are marked by red rings. On the right: an
    anti-aligned pair of the $j$th and the $k$th MTs.}
  \label{coarse}
\end{figure*}

To coarsegrain the BD-kMC simulation results and make connections with
the kinetic model, we first derive a continuum-mechanics model to
describe the MT dynamics. Here we assume the motor run length to be
approximately the MT length, meaning that once bound, the motors will
stay on the MTs until reaching the plus ends.  As shown in
Fig.~\ref{coarse}, we consider a nematically ordered local cluster of
MTs undergoing polarity sorting, with $n$ MTs pointing rightwards and
$m$ MTs pointing leftwards. Let all the MTs in this cluster be
coupled by active motors which create spring-like forces between the
MTs, and whose bound ends move at a characteristic (constant) speed
$v_w$ toward MT plus-ends. For an anti-polar MT pair this induces a
relative sliding, each towards its negative end. The cluster is
assumed small enough so that all MTs experience the same local flow
field. Using Stokesian slender body theory \cite{keller76} we can find
the velocities of the left- and rightward pointing MTs. For each MT,
the center locates at $x^c$, with the director ${\bf{p}}$. We assume
that in the cluster there are $m$ MTs pointing leftwards
(${\bf{p}} = -\mathbf{\hat{x}}$, with superscript $L$), and $n$ MTs
pointing rightwards (${\bf{p}} = \mathbf{\hat{x}}$, with superscript
$R$). Each anti-aligned pair (say the $j$th and the $k$th MT) shares
$Q$ ($Q > 1$) motors \beq {\bf{x}}_j^L = {\bf{x}}_j^{L,c} +
s_q^L{{\bf{p}}_j} = \left(x_j^{L,c} - s_q^L\right){\bf{\hat x}},\quad
{\bf{x}}_k^R = {\bf{x}}_k^{R,c} + s_q^R{{\bf{p}}_k} = \left(x_k^{R,c}
  + s_q^{R}\right){\bf{\hat x}}, \eeq where $j = 1..m, k = 1..n$ and
$q = 1..Q$.  As shown on the right in Fig.~\ref{coarse}, one motor
locates at $s_q^L\left( t \right) = s_{j,q}^{L,0} + {v_w}t$, and the
other locates at $s_q^R\left( t \right) = s_{k,q}^{R,0} + {v_w}t$,
with initial positions $s_{j,q}^{L,0}$ and $s_{k,q}^{R,0}$. The
characteristic motor speed $v_w$ is constant for the anti-aligned
pair. Hence the distance between the two motors in the tangential
direction can be calculated as %
\beq \Delta _{jk}^q{\bf{\hat x}} = {\bf{x}}_j^{L} - {\bf{x}}_k^{R} =
\left( {x_j^{L,c} - x_k^{R,c}} \right){\bf{\hat x}} - \left(
  {s_{j,q}^{L,0} + s_{k,q}^{R,0}} \right){\bf{\hat x}} - \left(
  {2{v_w}t} \right){\bf{\hat x}} = \left( {\Delta _{jk}^c - \Delta
    _{jk}^{q,0} - 2{v_w}t} \right){\bf{\hat x}}
\label{delta}
\eeq
where $\Delta _{jk}^c =-\Delta _{kj}^c= x_j^{L,c} - x_k^{R,c}$, $\Delta _{jk}^{q,0} = \Delta _{kj}^{q,0} = s_{j,q}^{L,0} + s_{k,q}^{R,0}$.
When the motor is walking, it behaves like a linear spring with rigidity $\kappa$ by exerting equal and opposite forces%
\beq
{\bf{f}}_{jk}^q =  - {\bf{f}}_{kj}^q =  - \kappa \Delta _{jk}^q{\bf{\hat x}}.
\label{spring}
\eeq As a result, the two MTs slide past one another undergoing
polarity sorting.  Following slender-body theory \cite{keller76}, the
MT speed is given by
${\bf{\dot x}}^c = \left( {\frac{{{\bf{I}} +
        {\bf{p}}{\bf{p}}}}{{\tilde \eta l }}} \right) \cdot
\sum\limits_{k = 1}^n {\sum\limits_{q = 1}^Q {{\bf{f}}} }$,
leading to \beq {\dot \Delta _{jk}^c} = \dot x_j^{L,c} - \dot
x_k^{R,c} = - \frac{{2\kappa }}{{\tilde \eta}l }\sum\limits_{k' = 1}^n
{\sum\limits_{q = 1}^Q {\left( {\Delta _{jk'}^c - \Delta _{jk'}^{q,0}
        - 2{v_w}t} \right)} } - \frac{{2\kappa }}{{\tilde\eta}l
}\sum\limits_{j' = 1}^m {\sum\limits_{q = 1}^Q {\left( {\Delta
        _{j'k}^c - \Delta _{j'k}^{q,0} - 2{v_w}t} \right)} }, \eeq
where $\tilde \eta = 4\pi \eta /\ln \left( 2r \right)$, and $\eta$ is
the fluid viscosity.  We seek the time-dependent solutions of the form
$\Delta_{jk}^{c}-\Delta_{jk}^{q,0}=A+B t$. The coefficients $A$ and
$B$ can be solved as \beq A = {\Delta _{jk}^q} = - \frac{{\tilde\eta l
    {v_w}}}{{Q\kappa \left( {m + n} \right)}},\quad B = 2{v_w},
\label{solution}
\eeq
leading to
\beq
{\dot x}_{j}^{L,c}  =\frac{n}{\left(  m+n\right)  }2v_w, \quad
{\dot x}_{k}^{R,c}  =-\frac{m}{\left(  m+n\right)  }2v_w,
\label{exact}
\eeq which suggests
$v^L=\frac{2n}{{n+m}}v_{w},\quad {v^R} = - \frac{2m}{n+m}v_{w}$. This
expression shows that the speed of each population depends on how many
opposing MTs there are to pull against, with their drag as the anchor,
and their relative velocity fixed at $v^L-v^R=2v_w$ by the motor
protein speed. This latter observation is in agreement with
observations of anti-aligned sliding of MTs in the mitotic spindle
\cite{yang08}.

Next, we consider a general situation when the MTs are not perfectly aligned but with an intersection angle, i.e., ${\bf p}_j\cdot {\bf p}_k = \pm 1 + O\left(\theta_{jk}^2\right)$ where $\theta_{jk}$ is a small angle between the $jth$ and the $kth$ MTs. As discussed later, at high concentration, the steric interactions align the neighbouring MTs, which makes the small-angle assumption a reasonable approximation. Similar to the perfectly-aligned case, the positions of the two motors can now be written as:
\beq
{{\bf{x}}_j} = {\bf{x}}_j^c + {s^q_j}{{\bf{p}}_j},\quad {{\bf{x}}_k} = {\bf{x}}_k^c + {s^q_k}{{\bf{p}}_k},
\eeq
where $j,k = 1..N, q = 1..Q$, and $s = {s^{q,0}} + {v_w}t$. So the relative distance becomes
\beq
{{\bf{\Delta }}_{jk}^q} = {\bf{x}}_j^c - {\bf{x}}_k^c + s_j^{q,0}{{\bf{p}}_j} - s_k^{q,0}{{\bf{p}}_k} + {v_w}t\left( {{{\bf{p}}_j} - {{\bf{p}}_k}} \right) = {\bf{\Delta }}_{jk}^c + {\bf{\Delta }}_{jk}^{q,0} + {v_w}t\left( {{{\bf{p}}_j} - {{\bf{p}}_k}} \right)
\eeq
where ${\bf{\Delta }}_{jk}^c = {\bf{x}}_j^c - {\bf{x}}_k^c$, ${\bf{\Delta }}_{jk}^{q,0} = s_j^{q,0}{{\bf{p}}_j} - s_k^{q,0}{{\bf{p}}_k}$. The motors exert tangential force ${{\bf{f}}_{jk}^q} =  - \kappa {{\bf{\Delta }}_{jk}}$.
Following the same procedure, we seek solutions of the form ${\bf{\Delta }}_{jk}^q = \left( {A + Bt} \right)\left( {{{\bf{p}}_j} - {{\bf{p}}_k}} \right)$, yielding
\beq
A = \frac{{\tilde\eta l {v_w}}}{{2QN\kappa }},\quad B =  - {v_w}.
\label{flux2}
\eeq
Then the relative moving speed of the two MTs becomes
\begin{eqnarray} {\bf{\dot \Delta }}_{jk}^c = {\bf{\dot x}}_j^c -
  {\bf{\dot x}}_k^c = - {v_w}\left( {{{\bf{p}}_j} - {{\bf{p}}_k}}
  \right).
\label{flux}
\end{eqnarray}
When ${{\bf{p}}_k} = -{{\bf{p}}_j}$, Eqs. (\ref{flux2}) and
(\ref{flux}) exactly recover the solutions in (\ref{exact}) for the
perfectly-aligned case. To further coarse grain the above results to
facilitate a continuum modeling as discussed below, we take an average
in $\bf p$ of Eq. (\ref{flux}) which directly yields a translational
particle flux ${\dot\bx} = \bq-\bp$.

\subsection{Flux velocity, active stress and kinetic model}
The system is described by a distribution function $\Psi(\bx,\bp,t)$
of MT center-of-mass positions $\bx$ and polar orientation vectors
$\bp$ ($|\bp|=1$), evolved through a Smoluchowski equation
\begin{equation}
\frac{{\partial \Psi }}{{\partial t}} + {\nabla _x} \cdot \left( {{\bf{\dot x}}\Psi } \right) + {\nabla _p} \cdot \left( {{\bf{\dot p}}\Psi } \right) = 0,
\label{smolu}
\end{equation}
which reflects conservation of particle number. Here ${\bf{\dot x}}$
and ${\bf{\dot p}}$ are MT conformational fluxes. Important
macroscopic quantities for describing a polar nematic system are the
local concentration $\Phi=\int_p\Psi$, the local polarity vector
$\bq=\int_p\Psi\bp/\Phi$, the second-moment tensor
$\bD=\int_p\Psi\bp\bp$ which arises generically in capturing
active stresses produced by active suspensions \cite{simha02}, the
(trace-free) order parameter tensor $\bQ=\bD/\Phi-\bI/d$, with $d=2$
or 3 the spatial dimension, and the fourth moment ${\bf S}= \int_p \Psi {{\bf{pppp}}}$.

Slender-body theory yields the forces each rod exerts on the
fluid, and hence the volume-averaged stress \cite{batchelor70} by
polarity sorting can be calculated. If the cluster occupies a volume
$V_c$, the induced extra stress tensor from anti-aligned sorting is
$\bSigma_{aa}=\frac{{\tilde\eta} v_wl^2}{V_c}
\frac{\alpha_{aa}}{2}\frac{2mn}{m + n} \bp\bp$. Here ${\tilde\eta}$ is
proportional to fluid viscosity $\eta$, and $\alpha_{aa}=s/l$ with $s$
the signed distance between the center-of-masses of the $\bp$ and
$-\bp$ oriented subclusters. If the cluster is extending then $s<0$,
as would be the case if motor protein binding and unbinding kinetics
biased motor densities towards the plus-end of the MTs. This is
seen in the BD-kMC simulations (Fig.~\ref{figure02}e), and is
associated with local extensile flows similar to those of motile
Pusher particles which collectively can drive macroscopic flow
instabilities\cite{saintillan08,saintillan08a,saintillan12}. The anti-aligned pair
stresslet strength can be derived as $S=\frac{{\tilde\eta} v_w l^2 \alpha_{aa}}{m+n}$.
When taking $v_w$ as $v_m$, we extract the value of $\alpha_{aa}\approx -2$
from the BD-kMC simulations.

While active motoring between polar-aligned MTs yields little MT
mobility, the BD-kMC simulations show that it does yield an extensile
stress. However, unlike polarity sorting we lack a simple
first-principles model of how polar interactions yield extensile
stress, though the number of polar pair interactions within a cluster
scales as $m^2+n^2$. Given that the anti- and polar-aligned stresses
are of the same order (Fig.~\ref{figure02}h) we assume the form
$\bSigma_{pa}=\frac{{\tilde\eta} v_wl^2}{V_c}
\frac{\alpha_{pa}}{2}\frac{m^2+n^2}{m+n}\bp\bp$. Comparison with the
BD-kMC simulations suggests that $\alpha_{pa}\approx -0.5$.

We have generalized this simple example to a continuum model that
captures polarity sorting of MTs and the dependence of the stress upon
the local polarity of the MT field.  The fluxes for Eq.~(\ref{smolu})
are given in dimensionless form by
\begin{eqnarray}
{\dot\bx} &=&(\bq-\bp)+\bU-D_t\nabla_x\ln \Psi,
\label{xflux} \\
{\dot\bp} &=& \left( \bI - \bp\bp\right)
\left( \nabla_x\bU + 2\zeta_0\bD \right) \bp - D_r\nabla_p\ln \Psi.
\label{pflux}
\end{eqnarray}
To non-dimensionalize the above equations, we assume that there are $M$ MTs in the entire computational domain of volume $V_c$.  At high concentration, it is useful to introduce an effective volume fraction $\nu = nbl ^2$ where $n = M/V_c$  is the mean number density \cite{doi88,ESS2013}. Further, we choose the characteristic length scale $l _c = b/\nu$, the velocity scale $v_c = v_{w}$, as well as the stress scale $\eta v_w/l _c$.
In Eq.~(\ref{xflux}), $\bU$ is the background fluid flow, and the last
term yields translational diffusion with constant $D_t$. For
nematically ordered suspensions, the term $\bq-\bp$ exactly reproduces
the cluster velocities induced by polarity sorting given above (note
that for a perfectly polar system, no polarity sorting occurs and the
flux $\bq-\bp$ makes no contribution.) In Eq.~(\ref{pflux}), the MTs
are rotated by the background flow gradient $\nabla_x \bU$ according
to Jeffery's equation \cite{Jeffery22} while the second term arises
from the Maier-Saupe potential with coefficient $\zeta_0$ which models
torques and stresses arising from steric interactions at high
concentration \cite{maier58,ESS2013}. The last term yields rotational
diffusion of the rod with constant $D_r$. We do not account for
MT rotation through interactions with the local field, as is
appropriate when the MT field is nematically ordered. All constants
have been made nondimensional using characteristic velocity $v_w$, and
a characteristic length $l_c$ appropriate for dense suspensions
\cite{doi88,ESS2013}.

Our system is closed by specifying how $\bU$ and $\nabla_x\bU$ are
recovered from $\Psi$, which involves specifying the extra stress
created in the fluid by activity and other sources. We assume the
active stress arises separately from anti-aligned and polar-aligned MT
interactions, and construct it from $\bD$ and $\Phi\bq\bq$ (i.e., the
simplest symmetric tensors quadratic in $\bp$).  In dimensionless
form, the active stress tensor takes the form
\begin{eqnarray}
  \bSigma^a= \frac{\alpha_{aa}}{2} ( \bD-\Phi\bq\bq )
+\frac{\alpha_{pa}}{2} ( \bD+\Phi\bq\bq ).
\label{extrastress}
\end{eqnarray}
The first term (second term) captures active stress production via
polarity sorting (motor relaxation) and exactly reproduces the form
of $\bSigma_{aa}$ ($\bSigma_{pa}$) for nematically ordered
suspensions.  The total extra stress tensor is given by
$\bSigma^e=\bSigma^a+\bSigma'$, where $\bSigma'$ models extra stresses
arising from flow-induced constraint forces on MTs and steric
interactions \cite{ESS2013}:
\beq
{\bSigma'} = \nu \beta {\bf S} :{\nabla _x}{\bf{U}} - 2\nu {\zeta _0}\beta \left( {\bf D}  \cdot {\bf D}  - {\bf S} :{\bf D}  \right)
\label{pstress2}
\eeq
where $\beta =\frac{{\pi }r}{{6\ln \left( {2r} \right)}}$.

For bulk flow modeling one typically closes the system by balancing
viscous and extra stresses and solving the forced Stokes equation
$-\nabla_xp+\Delta_x\bu=-\nabla_x\cdot\bSigma^e$ and
$\nabla_x\cdot\bu=0$ with velocity $\bu$ and pressure $p$. This
generates the background velocity and its gradient needed to evolve
Eq.~(\ref{smolu})\cite{saintillan08}. However, this
approach does not describe the streaming nematic experiments of
Sanchez {\it et al.} \cite{sanchez12}, where the active material is
confined to an interface between oil and water, so surface motions are
coupled to external fluid motions. To capture that coupling, we
consider a flat layer of interacting MTs bound in the $xy$-plane at
$z=0$, and immersed between two half-spaces filled with Newtonian
viscous fluid (for simplicity, of the same viscosity). The activity in
the MT layer generates a stress jump across the $z=0$ plane, and so
generates a global 3D flow which is continuous at $z=0$. In order to close the system, we solve the surface velocity ${\bf U}$ in terms of the extra stress ${\bf{\Sigma }}^e$. To accomplish this, we first solve the (3D) velocity field ${\bf u}=\left(u,v,w\right)$ of fluid flow using the Stokes equations
\beq
{\nabla} \cdot {\bf{u}} = 0,\quad  - \nabla^2{\bf{u}} + {\nabla}p = 0,
\eeq
where $\nabla$ is a regular 3D spatial gradient operator. Under
Fourier transform in
$(x,y)$, the above equations can be written as:
\beq
 - i{\bf{k}}\hat p + \left( {{\partial _{zz}} - {k^2}} \right){\bf{\hat v}} = 0,\quad  - {\hat p_z} + \left( {{\partial _{zz}} - {k^2}} \right)\hat w = 0,\quad i{\bf{k}} \cdot {\bf{\hat v}} + {\hat w_z} = 0,
\eeq
where ${\bf{k}}$ is a 2D wave-vector, and ${\bf{v}} =\left(
  u,v\right)$ is a 2D velocity field. When solving these equations in
the upper (+) and  lower (-) halves of the domain, we match at the MT
layer through the continuous (2D)
surface velocity ${\bf U}$, i.e., ${\bf{v}}^+ = {\bf{v}}^- = {\bf{U}}$
and $w^+ = w^- = 0$. After some algebra, we obtain
\beq
{{\hat p}_ \pm } =  - 2i{\bf{k}} \cdot {\bf{\hat U}}{e^{ \mp kz}},\quad {{{\bf{\hat v}}}_ \pm } = \left( {{\bf{I}} \mp kz{\bf{\hat k\hat k}}} \right){\bf{\hat U}}{e^{ \mp kz}},\quad {{\hat w}_ \pm } =  - i{\bf{k}} \cdot {\bf{\hat U}}z{e^{ \mp kz}},
\eeq
where $k = |{\bf{k}}|$ and ${\bf{\hat k}} = {\bf k}/k$ is a 2D unit wave-vector.
We further assume that the capillarity of the surface bounding the MT layer acts against concentration of MTs. We denote the liquid viscous stress as ${\bm{\sigma }} =  - p{\bf{I}} + \nabla {\bf{u}}$, and match the two solutions through a traction jump on the layer ${{\bm{\sigma }}^ + } \cdot {\bf{n}} - {{\bm{\sigma }}^ - } \cdot {\bf{n}} = \nabla_x  \cdot {{\bf{\Sigma }}}$. Here $\nabla_x$ is a 2D operator on the surface, and ${{\bf{\Sigma }}} = {\bf{\Sigma }}^e + {\bf{\Sigma }}^p$, arising from both the active stress ${\bf{\Sigma }}^e$ due to MT activity and the stress ${\bf{\Sigma }}^p$ due to a transverse pressure gradient within the MT layer which results in the background flow being incompressible in the plane (i.e., $\nabla_x \cdot {\bf{U}} = 0$). Then it is easy to eliminate ${\bf{\Sigma }}^p$ and solve the surface flow ${\bf U}$ in terms of ${\bf{\Sigma }}^e$ as:
\beq
{{\bf{\hat U}}_{{k}}} = \frac{i}{2}\left( {{\bf{I}} - {\bf{\hat
        k\hat k}}} \right) \left( {{{{\bf{\hat \Sigma }}}^e_{{k}}}
    {\bf{\hat k}}} \right).
\label{surfv}
\eeq
It is useful to compare this expression to that for the 2D Stokes
equation forced by a bulk stress: ${\hat\bu}=\frac{i}{k}(\bI -
{\hat\bk}{\hat\bk})({\hat\bSigma}^e{\hat\bk})$. The missing factor of
$k$ in Eq.~(\ref{surfv}) profoundly changes the nature of system
stability for the surface and 2D bulk systems. Equation
(\ref{surfv}) not only closes the system but facilitates a
pseudo-spectral method to solve the Smoluchowski equation
(\ref{smolu}) and the fluid flow in a coupled manner.

\subsection{Flow, polarity and defects}

\begin{figure*}
 \begin{center}
  \includegraphics[width = 164mm]{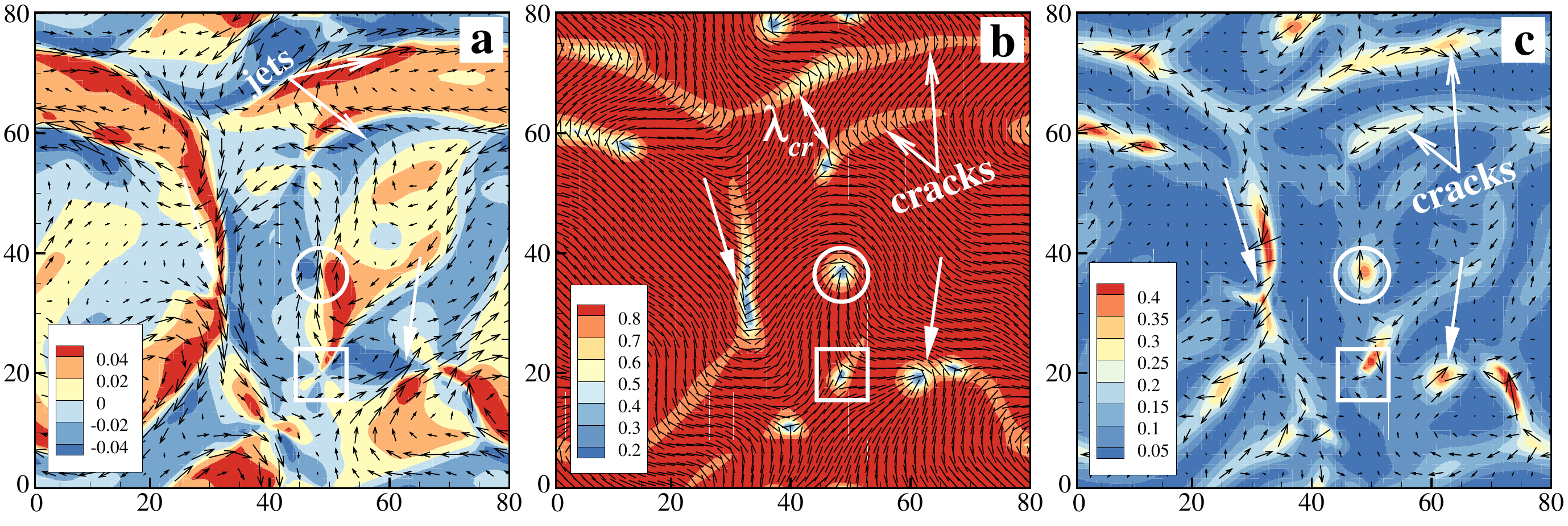}
 \end{center}
   \caption{Snapshots of streaming MT nematics on a liquid-liquid
    interface. The active stress magnitudes are chosen as $\alpha_{aa} = -2.0$ and
    $\alpha_{pa} = -0.6$. (a) The background fluid velocity vector field
    superposed upon the colormap of the associated vorticity. (b) The
    nematic director field $\bn$ superposed on the colormap of the
    scalar order parameter (twice the positive eigenvalue of the
    tensor $\bQ$). Disclination defects of order $\pm 1/2$ appear in
    localized regions of low order. Two defects are marked by an open
    circle ($+1/2$) and a square ($-1/2$). The arrow at right marks a
    pair of annihilating defects, while the arrow at left identifies
    an ``incipient crack'' from which a defect pair is about to emerge.
    Here $\lambda_{cr}$ is a calculated characteristic length between the cracks.
    Principal eigenvalues of the active stress due to polarity sorting (${\bf{\Sigma}}_{aa}$, c) and motor relaxation (${\bf{\Sigma}}_{pa}$, d).}
 \label{numerics}
\end{figure*}

Assuming 2D periodic boundary conditions, we have simulated our active
polar nematic model over long times, using
Eqs.~(\ref{smolu}-\ref{pstress2}) as well as the velocity-stress
kernel (\ref{surfv}). For the simulations shown here, we choose
$\alpha_{aa,pa}$ from $-0.1 \sim -4.0$, and fix $\beta = 1.74$ (i.e.,
aspect ratio 10), $\nu = 0.5$, $\zeta_0 = 1.0$, $D_t = 0.5$ and
$D_r = 0.1$ (estimated from the BD-kMC parameters). The computation is
performed on a 2D periodic domain of a square box with dimension
$L = 80$. The governing equations are solved spectrally in a coupled
manner, using the fast Fourier transform algorithm by expanding the
variables in Fourier series and truncating the series after 200 -- 400
modes in each spatial direction
\cite{saintillan07,saintillan08,ESS2013}.

Simulating in regions of flow instability we find persistently
unsteady flows correlated with continual genesis, propagation, and
annihilation of $\pm 1/2$ defect pairs.  When we examine simulation
results at late times, from initial data near uniform isotropy, we
find dynamics that are complex and appear turbulent, qualitatively
similar to those reported by Sanchez {\it et al.}  \cite{sanchez12}
(Fig.~\ref{numerics}). The surface velocity and vorticity show
formation of jets and swirls (Fig.~\ref{numerics}a, also see video S2). The local MT
orientation is highly correlated with the flow structures, and the
surface is littered with $\pm 1/2$ defects which propagate freely
about the system (Fig.~\ref{numerics}b, also see video S3).  These defects exist in
regions of small nematic order (dark blue), and are born as opposing
pairs in elongated ``incipient crack'' regions. These are associated
with surface jets, locally decreasing nematic order, and increasing
curvature of director field lines. Characteristically, the $+1/2$
defects propagate away along their central axis and have a much higher
velocity than those of $-1/2$ order. The relatively higher surface
velocity in the neighborhood of a $+1/2$ defect appears as a
well-localized jet, in the direction of defect motion, between two
oppositely signed vortices.

The active force vector field $\bff^a=\nabla_x\cdot\bSigma^a$ is
correlated with regions of rapidly changing nematic order
(Fig.~\ref{numerics}c, also see video S4).  Large active force is present along an
interconnected network of ridges correlated with the stringy regions
of diminished nematic order and particularly with incipient
cracks. Along such cracks, the active force points in the direction
from which newly nucleated $+1/2$ defects will emerge and propagate.
Isolated high-force peaks correlate and move with $+1/2$ defects, with
the force pointing in the direction of their motion. Negative order
defects are associated with regions of relatively low force magnitude,
likely due to the local symmetry of the nematic director field.

\begin{figure*}
 \begin{center}
  \includegraphics[scale = 0.6]{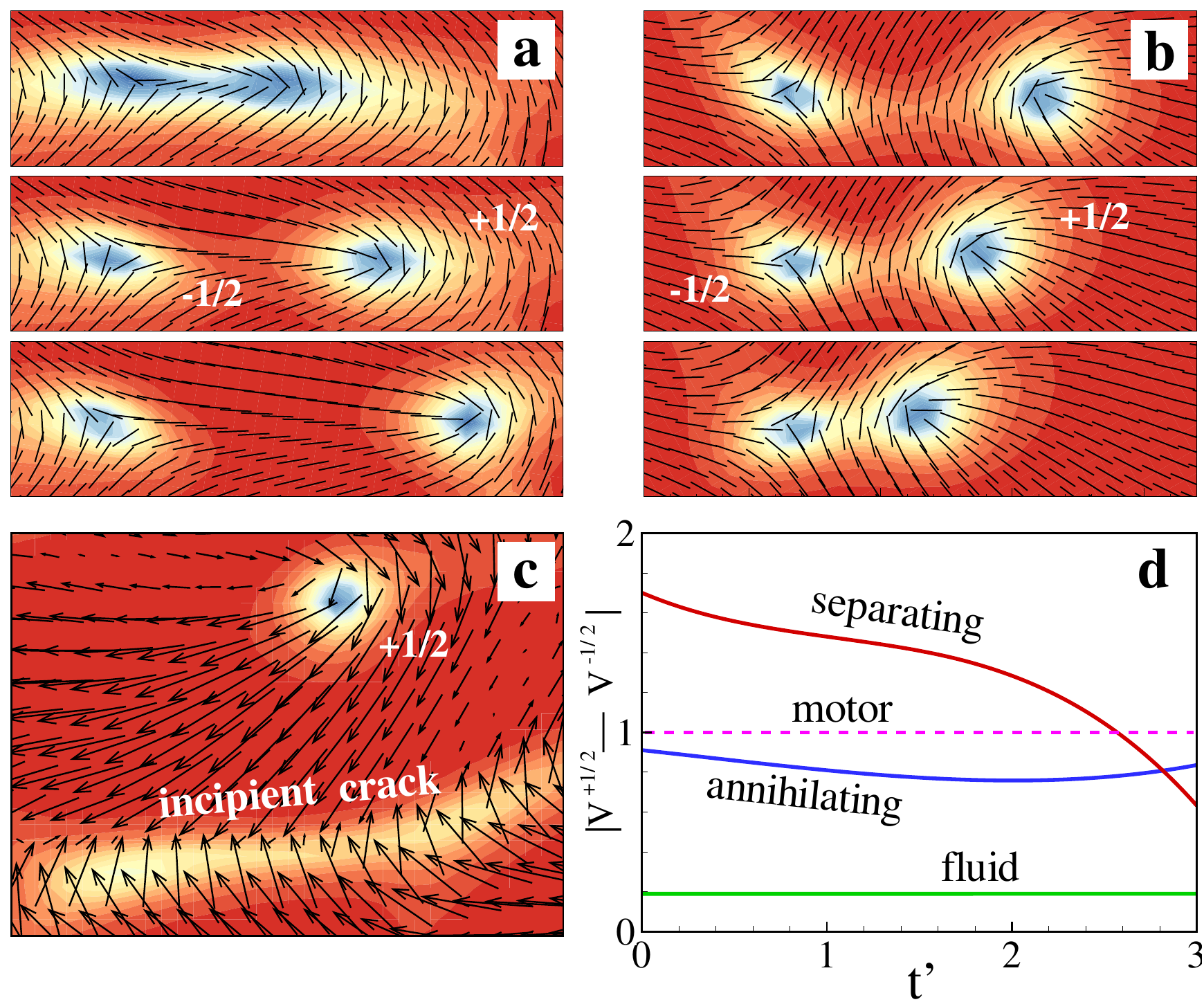}
 \end{center}
 \caption{Time sequential snapshots of the nematic director field
   $\bn$ for nucleation (a) and annihilation (b) of defect pairs,
   where $\alpha_{aa} = -2.0$ and $\alpha_{pa} = -0.6$ are fixed. The
   (dimensionless) time spacing between frames is 5.  (c) Polarity
   field associated with a motile $+1/2$ defect and an incipient crack
   on the bottom.  (d) Relative speed of the two oppositely charged
   defects, as well as the mean flow speed near this defect pair, as function of dimensionless time $t'$.  In
   (a), (b) and (c), the color shows the scalar order parameter,
   plotted with the same scale as Fig.  \ref{numerics}b. }
\label{interface}
\end{figure*}

\begin{figure*}[t!]
  \centering
  \includegraphics[scale = 0.57]{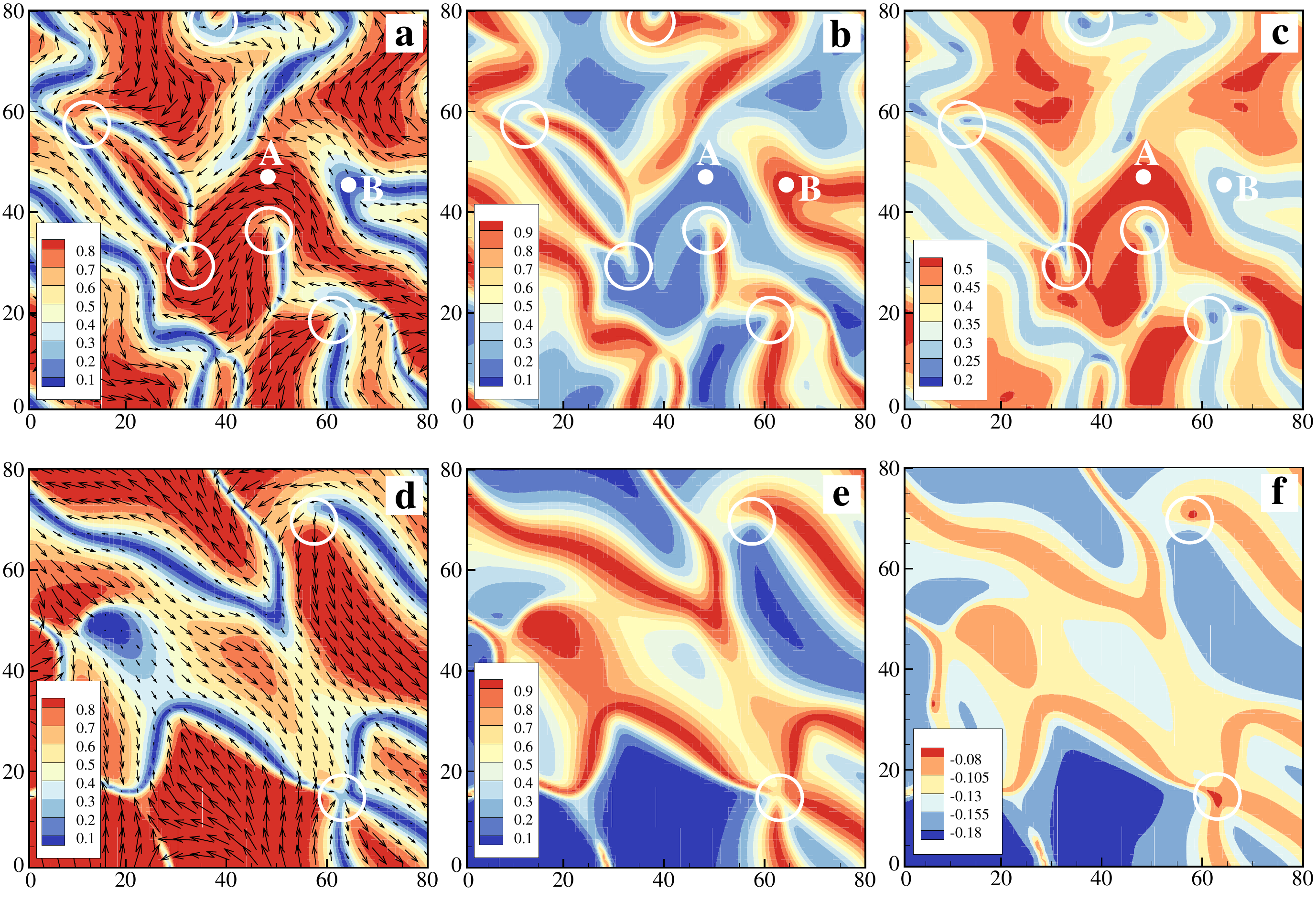}
  \includegraphics[scale = 0.57]{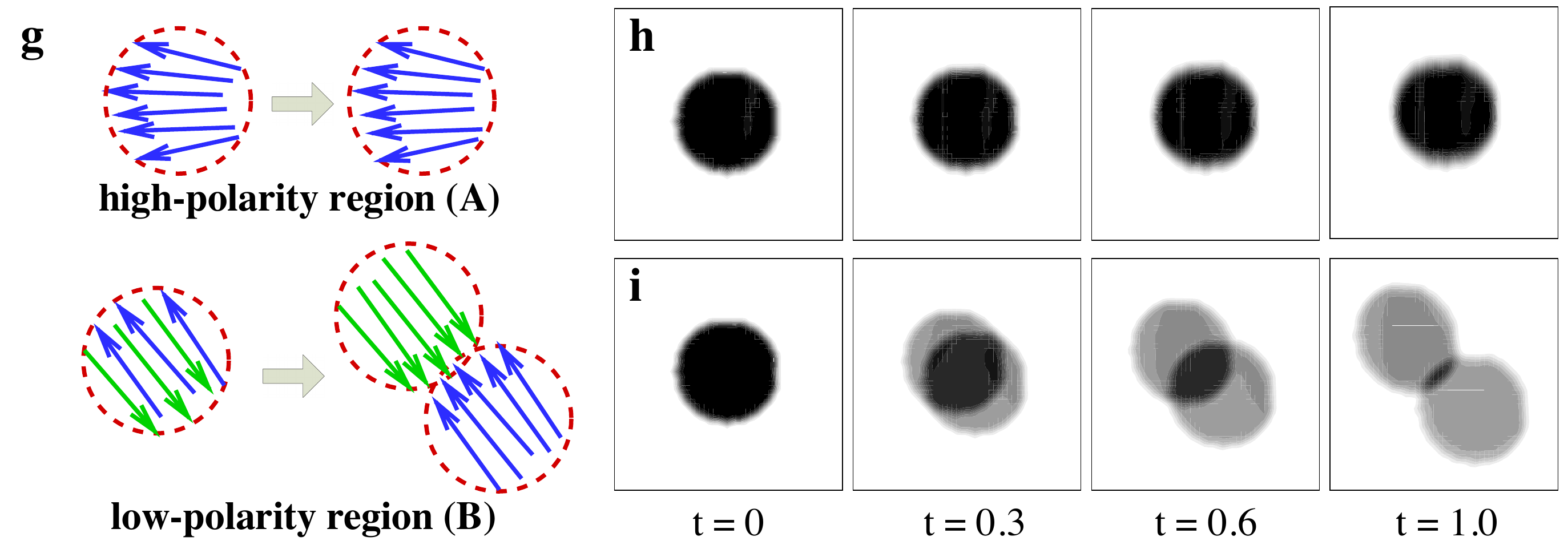}
  \caption{Dynamics of the polarity field, the polarity-dependent
    active stresses, and the predicted dynamics of a photobleaching
    experiment. (a-c) Results from the simulation shown in
    Fig.~\ref{numerics}. (a) The polarity vector field $\bq$
    superimposed upon its magnitude $|\bq|$ (the local polar
    order). Circular areas labeled A and B mark regions of high and
    low polarity, respectively. (b,c) Polarity-dependent active stress
    magnitudes, showing principal eigenvalues of the active stresses
    due to polarity sorting (${\bf{\Sigma}}_{aa}$, b) and motor
    relaxation (${\bf{\Sigma}}_{pa}$, c). In (a-c), the stress
    magnitudes are chosen as $\alpha_{aa} = -2.0$ and
    $\alpha_{pa} = -0.6$. For comparison, (d-f) shows the polarity
    field and the polarity-dependent active stress fields when
    choosing $\alpha_{aa} = -2.0$ and $\alpha_{pa} = 0.2$. In (a-f),
    positions of $+1/2$-order defects are marked by open circles. (g)
    Schematic of predicted dynamics for a bleached spot of high
    nematic order in a region of high polar order (area A), and in a
    region of low polar order (area B). Arrows represent MTs with
    arrowheads denoting plus ends. In panels (h) and (i) these
    predictions are borne out by simulations of photobleached spots in
    areas A and B, respectively.}
  \label{polarity}
\end{figure*}

We observe both nucleation and annihilation of defect pairs
(Fig.~\ref{interface}). The birth and separation of a defect pair
begins from an incipient crack wherein the initially smooth director
field (e.g., lower arrow in Fig.~\ref{numerics}b) morphs into singular
forms in regions of low nematic order
(Fig.~\ref{interface}a). Typically the positively-signed defect moves
away faster and roughly along its symmetry axis. Following
annihilation of an oppositely-charged defect pair
(Fig.~\ref{interface}b), the nematic order increases as the director
field reknits itself into a smooth form (e.g., upper arrow in
Fig.~\ref{numerics}b). We examined how the polarity field $\bq$
changes near a defect and incipient crack (Fig.~\ref{interface}c).  As
the $+1/2$ defect propagates, it leaves behind a region of increased
polarity.  The polarity field rapidly rotates across the incipient
crack (by approximately $\pi/2$), and sometimes forms a shock-like
structure that precedes the birth of a new defect pair. We measured
the relative speed of the defect pairs (Fig.~\ref{interface}d).  The
speeds are similar to each other and on the order of the motor protein
speed in our model (normalized to unity).  This is consistent with
experimental observations ({\it cf.}  Fig.~3 of Sanchez {\it et al.}
\cite{sanchez12}).  The average fluid velocity around the defect pair
is much lower than the defect speeds. Hence, as is the case for
defects in more standard liquid crystalline materials, the defects
here are not material structures carried along by the background
surface flow.

Because our model is based on polar-specific fluxes and active
stresses, the polarity field $\bq$ \cite{Tjhung11}, polarity-dependent
active stresses $\bSigma_{aa}$ and $\bSigma_{pa}$, and the local MT
dynamics are coupled (Fig.~\ref{polarity}).  The polarity field
develops considerable spatial variation with regions of high and low
polar order (Fig.~\ref{polarity}a, also see video S5).  The two active stresses vary in
strength depending on the local polarity -- the polar-aligned
(anti-aligned) stress is large in regions of high (low) polar order
(Fig.~\ref{polarity}b,c). The anti-aligned stress yields the largest
forces, by about a factor of 3 (close to the ratio
$\alpha_{aa}/\alpha_{pa}$). The polarity field varies rapidly around
$+1/2$ defects, leading to gradients in the active stresses and large
active force (open circles in Fig.~\ref{polarity}). For comparison, we
did another numerical test where we assume the active stress generated
during motor tether relaxation is contractile (i.e., $\alpha_{pa}>0$,
Fig.~\ref{polarity}d-f). The ratio between the anti-aligned and
polar-aligned stress is still close to the ratio
$|\alpha_{aa}|/|\alpha_{pa}|$.  However, since the sign of the
polar-aligned stress changes, the two stresses exist in approximately
the same regions.

To illustrate the dramatic variation of local MT fluxes with the local
polarity field, we simulated the results of a photobleaching
experiment in which a circular region is exposed to high-intensity
laser light to bleach the fluorescing molecules on the corresponding
MTs \cite{AKSEW1976} (Figs.~\ref{polarity}g-i). In a small
high-polarity region (marked A in Fig.~\ref{polarity}a), little or no
polarity sorting occurs. Therefore the photobleached spot remains
approximately circular (Fig.~\ref{polarity}g top, h) and would deform
due to the fluid flow over longer times. In a low-polarity region of
high nematic order (marked B in Fig.~\ref{polarity}b), strong polarity
sorting of anti-aligned MTs causes a photobleached spot to separate
into two lobes (Fig.~\ref{polarity}g lower, i). Each lobe mixes with
unbleached surrounding MTs due to their active relative flux, showing
decreased bleaching. Through the lens of our theory, this type of
experiment probes the local polarity field, and hence the origins of
active stress.

\subsection{Coherent structures and hydrodynamic instabilities}
\label{hydroinstability}
 In our simulations, defect pairs are generated
 along elongated cracks that develop in regions of high
 polar order. To understand this instability, we consider
 nematically-aligned MTs using reduced equations where
particle diffusion is neglected (i.e., $D_r = D_t = 0$) in (\ref{pflux}).
We then adopt bipolar solutions of the
form $\Psi(x,p,t) = A(x,t)\delta(\bp-\bq_1(x,t)) + B(x,t)\delta(\bp-
\bq_2(x,t))$, where the concentrations $A$ and $B$ and orientations
$\bq_{1,2}$ are governed by:
\beq
\left\{ \begin{array}{l}
\frac{\partial A}{\partial t} + {\nabla _x} \cdot \left( {A{\bf{U}}} \right) - {\nabla _x} \cdot \left[ {\frac{{AB}}{{A + B}}\left( {{{\bf{q}}_1} - {{\bf{q}}_2}} \right)} \right] = 0,\\
\frac{\partial B}{\partial t} + {\nabla _x} \cdot \left( {B{\bf{U}}} \right) + {\nabla _x} \cdot \left[ {\frac{{AB}}{{A + B}}\left( {{{\bf{q}}_1} - {{\bf{q}}_2}} \right)} \right] = 0,\\
\frac{\partial {\bf{q}}_{1}}{\partial t} + {\bf{U}} \cdot {\nabla _x}{{\bf{q}}_1} - \frac{B}{{A + B}}\left( {{{\bf{q}}_1} - {{\bf{q}}_2}} \right) \cdot {\nabla _x}{{\bf{q}}_1} = \left( {{\bf{I}} - {{\bf{q}}_1}{{\bf{q}}_1}} \right) \cdot {\nabla _x}{\bf{U}} \cdot {{\bf{q}}_1} + 2{\zeta _0}B\left( {{{\bf{q}}_1} \cdot {{\bf{q}}_2}} \right)\left( {{\bf{I}} - {{\bf{q}}_1}{{\bf{q}}_1}} \right) \cdot {{\bf{q}}_2}, \\
\frac{\partial {\bf{q}}_{2}}{\partial t} + {\bf{U}} \cdot {\nabla _x}{{\bf{q}}_2} + \frac{A}{{A + B}}\left( {{{\bf{q}}_1} - {{\bf{q}}_2}} \right) \cdot {\nabla _x}{{\bf{q}}_2} = \left( {{\bf{I}} - {{\bf{q}}_2}{{\bf{q}}_2}} \right) \cdot {\nabla _x}{\bf{U}} \cdot {{\bf{q}}_2} + 2{\zeta _0}A\left( {{{\bf{q}}_1} \cdot {{\bf{q}}_2}} \right)\left( {{\bf{I}} - {{\bf{q}}_2}{{\bf{q}}_2}} \right) \cdot {{\bf{q}}_1}.
\end{array} \right.
\label{align}
\eeq
We consider the solutions for two groups of MTs undergoing polarity sorting along $\hat{\bf{x}}$: $A = \frac{1}{2} + \varepsilon A',\quad B = \frac{1}{2} + \varepsilon B',\quad {{\bf{q}}_1} = {\bf{x}} + \varepsilon {{\bf{q'}}_1},\quad {{\bf{q}}_2} =  - {\bf{x}} + \varepsilon {{\bf{q'}}_2}, \quad \bf{U} = \varepsilon \bf{u'}$, when $\bq_{1,2}\cdot {\hat{\bf x}} = 0$ which ensures that the length of $\bq_{1,2}$ remains 1 to order $\varepsilon^2$ for $|\varepsilon|  \ll 1$.  At order $\varepsilon$, we obtain a set of linearized reduced equations:
\beq
\left\{ \begin{array}{l}
\frac{\partial \left( {A' + B'} \right) }{{\partial t}} + \nabla_x  \cdot {\bf{u'}} = 0,\\
\frac{{\partial \left( {A' - B'} \right)}}{{\partial t}} - \frac{1}{2}\nabla_x  \cdot \left( {{{{\bf{q'}}}_1} - {{{\bf{q'}}}_2}} \right) - {\bf{\hat x}} \cdot \nabla_x \left( {A' + B'} \right) = 0,\\
\frac{{\partial \left( {{{{\bf{q'}}}_1} + {{{\bf{q'}}}_2}} \right)}}{{\partial t}} - {\bf{\hat x}} \cdot \nabla_x \left( {{{{\bf{q'}}}_1} - {{{\bf{q'}}}_2}} \right) =  - 2{\zeta _0}\left( {{{{\bf{q'}}}_1} + {{{\bf{q'}}}_2}} \right), \\
\frac{{\partial \left( {{{{\bf{q'}}}_1} - {{{\bf{q'}}}_2}} \right)}}{{\partial t}} - {\bf{\hat x}} \cdot \nabla_x \left( {{{{\bf{q'}}}_1} + {{{\bf{q'}}}_2}} \right) = 2\left( {1 - {\bf{\hat x\hat x}}} \right) \cdot \nabla_x {\bf{u'}} \cdot {\bf{\hat x}},
\end{array} \right.
\eeq
and the linearized active stress
${{\bf{\Sigma }}^e} = \left( {{\alpha _1} + {\alpha _2}} \right){\bf{\hat x\hat x}} + \varepsilon \left( {\frac{{{\alpha _1} + {\alpha _2}}}{2}} \right)\left[ {\left( {{{{\bf{q'}}}_1} - {{{\bf{q'}}}_2}} \right){\bf{\hat x}} + {\bf{\hat x}}\left( {{{{\bf{q'}}}_1} - {{{\bf{q'}}}_2}} \right)} \right] + \varepsilon \nu \beta \left( {{\bf{\hat x\hat x\hat x\hat x}}} \right):\nabla_x {\bf{u'}}$ in the
velocity-stress kernel in Eq.(\ref{surfv}). Next, we consider
plane-wave solutions $\Psi '\left( {{\bf{x}},{\bf{p}},t} \right) =
\tilde \Psi \left( {{\bf{k}},{\bf{p}}} \right)\exp \left( {i{\bf{k}}
    \cdot {\bf{x}} + \sigma t} \right)$ and ${\bf{u'}}\left(
  {{\bf{x}},t} \right) = {\bf{\tilde u}}\left( {\bf{k}} \right)\exp
\left( {i{\bf{k}} \cdot {\bf{x}} + \sigma t} \right)$, and assume that
$\hat{\bf k}$ lies in the plane defined by $\left(\hat{\bf x},\bq'_1 -
  \bq'_2\right)$.  The dispersion
relation can be solved analytically with the two branches of
solutions:
\beq
{\sigma _{1,2}}
= \frac{{f\left( \theta \right)}}{2} - {\zeta_0} \pm \sqrt {{{\left[
    {{\zeta_0} + \frac{{f\left( \theta \right)}}{2}} \right]}^2} -
 \frac{{{k^2}}\cos {{\left( \theta \right)}^2}}{4}}
\eeq
where
\beq
f\left( \theta  \right) =  - \frac{{\left( {{\alpha _1} + {\alpha _2}} \right)k\cos {{\left( \theta  \right)}^2}\cos \left( {2\theta } \right)}}{{2 + \nu \beta k\cos {{\left( \theta  \right)^2}}\cos {{\left( 2 \theta  \right)}}}}.
\label{ftheta}
\eeq
As $k \rightarrow 0$, the growth rate approaches two solutions: $\sigma_1 = f(\theta)$ and $\sigma_2 = -2\zeta_0$, which clearly illustrates a competition between a destabilizing effect due to the active stress and a stabilizing effect due to MT alignment through steric interactions. At large $k$, the growth rate has an asymptotic limit ${\mathop{\rm Re}\nolimits} \left( \sigma  \right) \rightarrow {{f\left( \theta  \right)}}/{2} - {\zeta_0}$. In addition, we find that the fluid constraining stress tends to stabilize the system by effectively decreasing the magnitude of the active stress in $f\left(\theta\right)$.

\begin{figure*}
 \begin{center}
  \includegraphics[scale = 0.6]{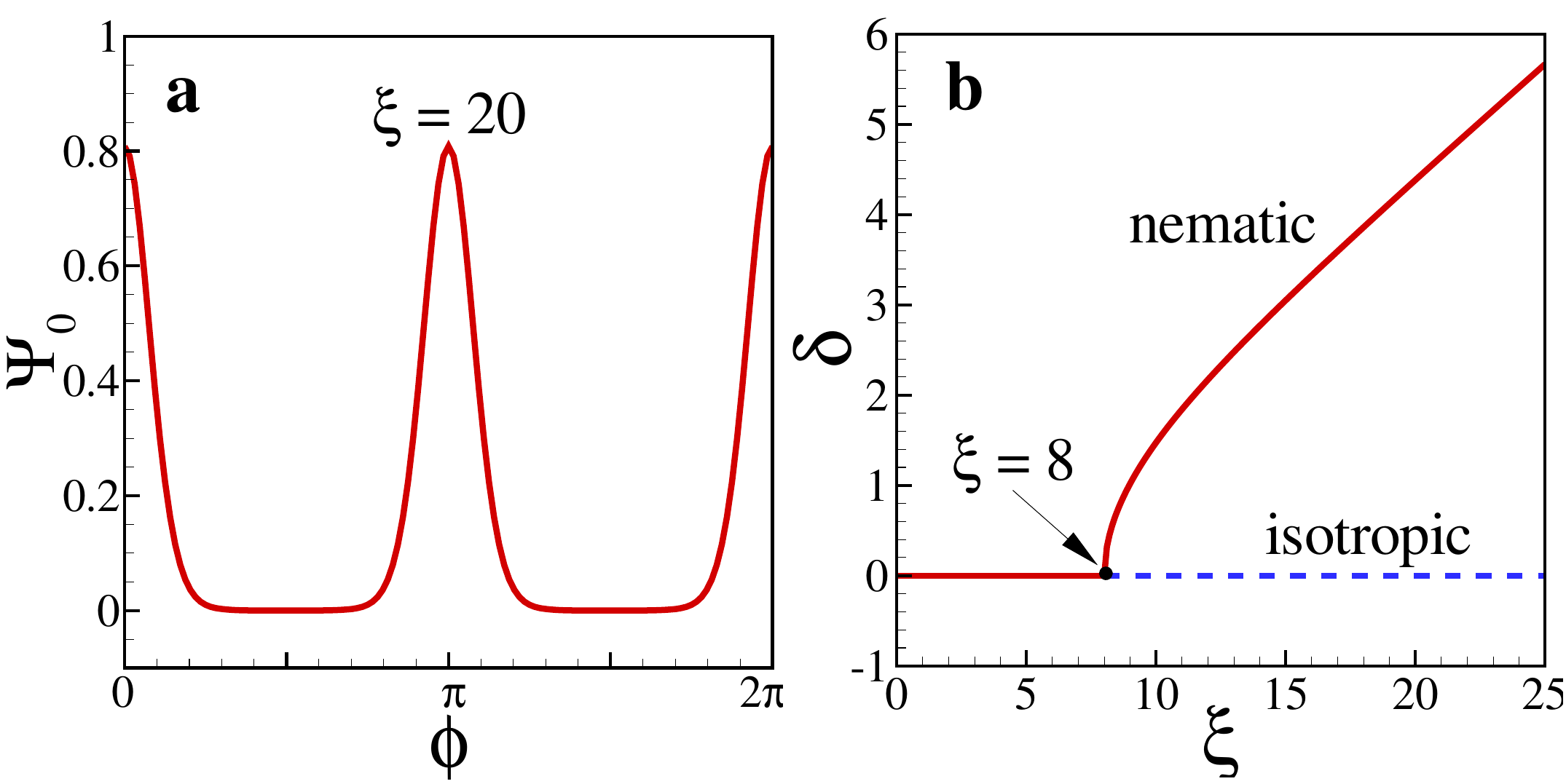}
 \end{center}
 \caption{(a) Steady-state solution $\Psi_0$ as a function of the orientation angle $\phi$ when choosing $\delta = 4.38$ and $\xi = 20$. (b) $\delta$ as a function of $\xi$. The bifurcation occurs at $\xi=8$.}
 \label{psi}
\end{figure*}

Next, we perform linear stability analysis for full nonlinear equations. At the nematically ordered base state, we seek a spatially uniform solution in 2D by balancing the angular diffusion and the alignment torque as a result of steric interactions in the rotational flux \cite{ESS2013}, i.e.,
\beq
{\nabla _p}\ln {\Psi _0} = \xi \left( {{\bf{I}} - {\bf{pp}}} \right) {{\bf{D}}_0} {\bf{p}}
\eeq
where $\xi = 2\zeta _0/D_r$. This equation admits a symmetric solution
as shown in Fig. \ref{psi}a:
\beq
{\Psi _0} = \frac{{\exp \left[ {\delta \left( \xi  \right)\cos \left( {2\phi } \right)} \right]}}{{\int {d\phi '} \exp \left[ {\delta \left( \xi  \right)\cos \left( {2\phi '} \right)} \right]}},
\label{alignedsol}
\eeq
where $\delta$ satisfies
\beq
\delta  = \frac{\xi }{4}\frac{{\int {d\phi '\cos \left( {2\phi '} \right)\exp \left[ {\delta \cos \left( {2\phi '} \right)} \right]} }}{{\int {d\phi '\exp \left[ {\delta \cos \left( {2\phi '} \right)} \right]} }}.
\label{dlta}
\eeq
We then perform a shift in coordinates and rewrite the equation as
\beq
g(\delta) = \delta  - \frac{\xi }{4}I\left( \delta  \right) = 0.
\eeq
We numerically calculate $\delta$ as a function of $\xi$. For small
$\xi$ there is only one solution, $\delta = 0$  associated with
$\Psi_0 =
\frac{1}{2\pi}$. This bifurcates into two solutions when
$g'(0) = 0$. Therefore, we have
\beq
g'\left( 0 \right) = 1 - \frac{\xi }{4}\frac{d}{{d\delta }}{\left. {\frac{{\int_0^{2\pi } {\sin \left( \omega  \right)\exp \left[ {\delta \sin \left( \omega  \right)} \right]d\omega } }}{{\int_0^{2\pi } {\exp \left[ {\delta \sin \left( \omega  \right)} \right]d\omega } }}} \right|_{\delta  = 0}} = 1 - \frac{\xi }{8},
\eeq
which gives that there is a second solution only for $\xi>8$ (see Fig. \ref{psi}b), suggesting that
in two-dimensions, the
Maier-Saupe potential yields an isotropic to nematic phase transition,
with increasing $\zeta_0$, when $\zeta_0=4D_r$. For all the simulations shown in the paper, we fix $\zeta _0 = 1.0$ and $D_r = 0.1$, which corresponds to $\xi = 20$ and $\delta = 4.38$.

We perturb the nematically-ordered base-state solution such that $\Psi  = {\Psi _0}\left( {\bf{p}}  \right) + \varepsilon \Psi '\left( {{\bf{x}},{\bf{p}} ,t} \right), {\bf{U}}\left( {\bf{x}} \right) = \varepsilon {\bf{u}} '\left( {\bf{x}} \right)$, leading to a linearized Smoluchowski equation for $\Psi '$:
\beq
{{\Psi '}_t} + {\Psi _0}{\nabla _x} \cdot {\bf{q'}} - {\bf{p}} \cdot {\nabla _x}\Psi ' + {\nabla _p} \cdot \left[ {{\Psi _0}\left( {{\bf{I}} - {\bf{pp}}} \right) \left( {{\nabla _x}{\bf{u'}} + 2{\zeta _0}{\bf{D'}}} \right) {\bf{p}} + 2{\zeta _0}\left( {{\bf{I}} - {\bf{pp}}} \right){{\bf{D}}_0}{\bf{p}}\Psi '} \right] = {D_t}{\Delta _x}\Psi ' + {D_r}{\Delta _p}\Psi '.
\eeq
By using the plane-wave solutions $\Psi '\left( {{\bf{x}},{\bf{p}},t}
\right) = \tilde \Psi \left( {{\bf{k}},{\bf{p}}} \right)\exp \left(
  {i{\bf{k}} \cdot {\bf{x}} + \sigma t} \right)$ and ${\bf{u'}}\left(
  {{\bf{x}},t} \right) = {\bf{\tilde u}}\left( {\bf{k}} \right)\exp
\left( {i{\bf{k}} \cdot {\bf{x}} + \sigma t} \right)$, this can be
rewritten as:
\beq
\sigma \tilde \Psi  + {\Psi _0}\left( {i{\bf{k}} \cdot {\bf{\tilde q}}} \right) - i\left( {{\bf{p}} \cdot {\bf{k}}} \right)\tilde \Psi
+ {\nabla _p} \cdot \left[ {{\Psi _0}\left( {{\bf{I}} - {\bf{pp}}} \right) \left( {i{\bf{\tilde uk}} + 2{\zeta_0}{\bf{\tilde D}}} \right) {\bf{p}} + 2{\zeta_0}\left( {{\bf{I}} - {\bf{pp}}} \right) {{\bf{D}}_0}  {\bf{p}}\tilde \Psi } \right]
=  - {D_t}{k^2}\tilde \Psi  + {D_r}{\Delta _p}\tilde \Psi.
\label{linearSmo}
\eeq
The perturbed velocity field satisfies
\beq
{\bf{\tilde u}} = \frac{i}{2}\left( {{\bf{I}} - {\bf{\hat k\hat k}}}
\right){{\bf{\tilde \Sigma}}^e{ \bf {\hat k}}},
\label{kernel}
\eeq
with the linearized extra stress
\beq
{\bf{\tilde \Sigma }}^e = \left( {{\alpha _{aa}} + {\alpha _{pa}}} \right){\bf{\tilde D}} + \nu \beta {{\bf{S}}_0}:\left( {i{\bf{\tilde uk}}} \right) - 2\nu {\zeta_0}\beta \left( {{{\bf{D}}_0} \cdot {\bf{\tilde D}} + {\bf{\tilde D}} \cdot {{\bf{D}}_0} - {{\bf{S}}_0}:{\bf{\tilde D}} - {\bf{\tilde S}}:{{\bf{D}}_0}} \right).
\label{linearStress}
\eeq
In the above equations, ${\bf p} =
\left[\cos\left(\phi\right),\sin\left(\phi\right)\right]$,
${{\bf{D}}_0} = \int_p {{\bf{pp}}{\Psi _0}} , {{\bf{S}}_0} = \int_p
{{\bf{pppp}}{\Psi _0}}, {\bf{\tilde q}} = \int_p {{\bf{p}}\tilde \Psi
}, {\bf{\tilde D}} = \int_p {{\bf{pp}}\tilde \Psi }$ and ${\bf{\tilde
    S}} = \int_p {{\bf{pppp}}\tilde \Psi }.$ By changing direction of
the wave-vector ${\bf{\hat k}}$, we discretize $\Psi'$ and use
pseudospectral collocation in the $\phi$ direction with 256 modes, and
numerically solve the eigenvalue problem for
Eqs. (\ref{linearSmo})-(\ref{linearStress}) to obtain the growth rate
\cite{saintillan08}.

\begin{figure*}
 \begin{center}
  \includegraphics[width = 164mm]{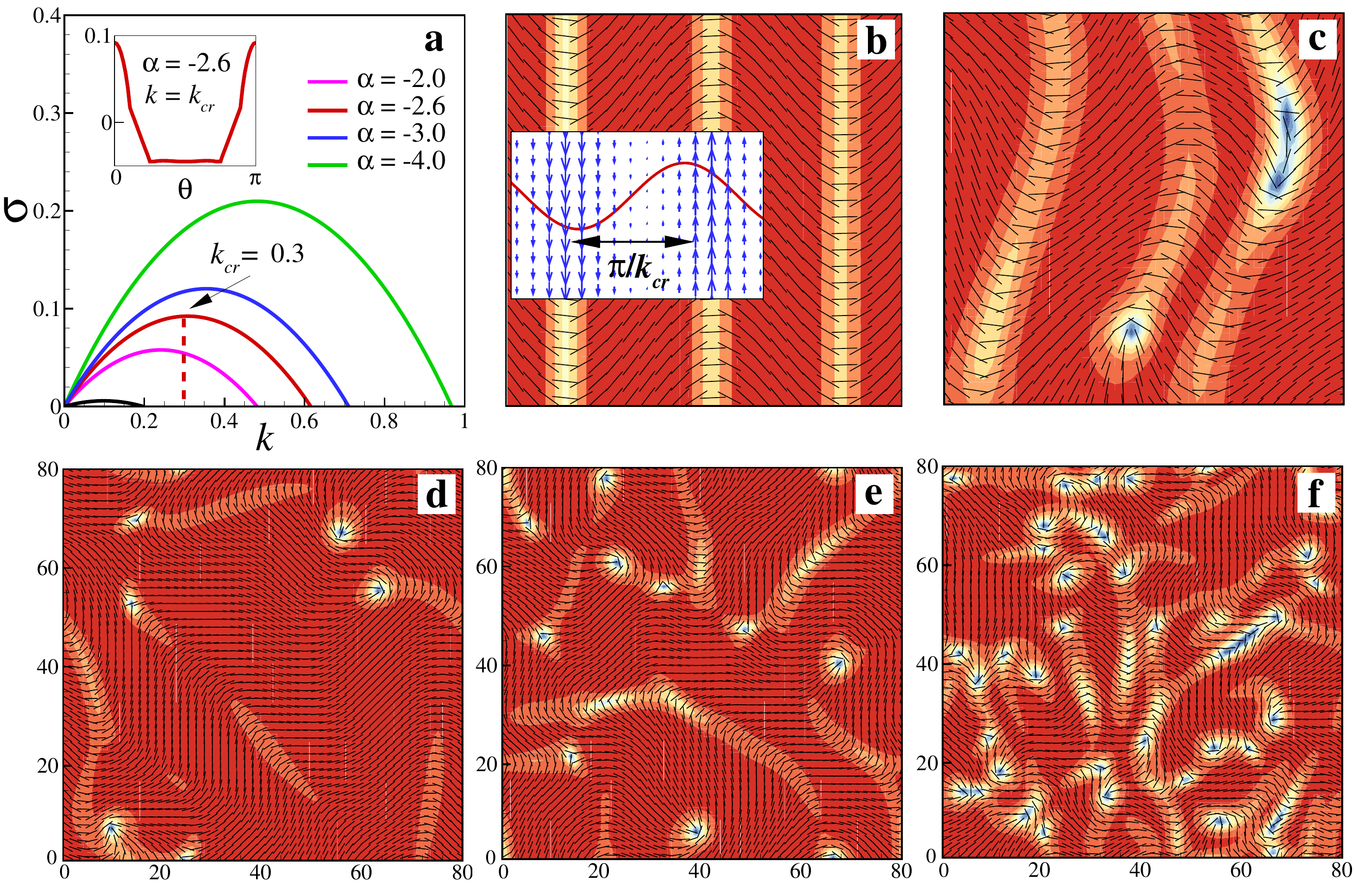}
 \end{center}
 \caption{ Linear stability analysis (a) and nonlinear simulation
   (b,c) for strongly anti-aligned MTs. (a) The real part of the
   growth rate as a function of wave-number $k$ for several $\alpha$
   ($\alpha = \alpha_{aa} + \alpha_{pa}$). $k_{cr}$ corresponds to the
   maximum growth rate. Inset: real part of the growth rate as a
   function of wave-angle $\theta$ when fixing $k = k_{cr}$. The black line represents the growth rate of the laning instability (without hydrodynamics). (b)
   Cracks formation. Inset: the fluid velocity vector field (blue) and
   the eigenmode (red line) associated with $k_{cr}$ for the linear system.
   (c) Genesis of defects at late times. In (d),(e) and (f), the active-stress magnitudes are chosen as $\alpha_{aa} = -0.4, -2.0$ and $-4.0$, respectively, while $\alpha_{pa} = -0.6$ is fixed. The snapshots accordingly show the distribution of
   disclination defects streaming in the domain with changing length scale.
   In (b-f), the nematic
   director field $\bn$ is plotted with the contour map of the
   scalar order parameter. The color scale is the same as Fig. \ref{numerics}b.}
\label{instability}
\end{figure*}

We find that the plane-wave vector of maximal growth is aligned with
the nematic director ($\theta=0$ in Fig.~\ref{instability}a inset;
also see \cite{saintillan08,saintillan08a,ESS2013}) with a wave-number
of maximal growth, $k_{cr}$, along this direction
(Fig.~\ref{instability}a). We find $k_{cr}$ grows approximately
linearly with $\alpha=\alpha_{aa}+\alpha_{pa}$. In the 2D bulk model,
the maximal growth occurs at $k=0$, and so does not produce a
characteristic length scale. However in this immersed layer system,
long-wave growth is cut off (see discussion following
Eq.~(\ref{surfv})) and yields a finite length scale of maximal growth.
Similar effects have been reported by Leoni and Liverpool
\cite{leoni10} in their study of swimmers confined to immersed thin
films, while Thampi {\it et al.} \cite{thampi14c} show that adding
substrate friction changes length-scale selection in 2D active nematic
models.

The result of this instability is captured in nonlinear simulations by
perturbing an MT suspension that is aligned along $\hat{\bf x}$,
causing a series of cracks to form along $\hat{\bf y}$
(Fig.~\ref{instability}b). These cracks are associated with up and
down moving fluid jets and bending of nematic field lines.  The
spatial variations of the velocity field are in excellent agreement
with the velocity eigenmode associated with $k_{cr}$ for the
linearized system (Fig.~\ref{instability}b inset): the distance
between these cracks matches the half-wavelength, i.e.,
$\lambda_{cr} = \pi \, /k_{cr}\approx 10$, which is in fact
representative of the characteristic length between cracks seen in the
full dynamics of motile defects (Fig. \ref{numerics}b). At late times,
these cracks lose stability when interacting with each other, and are
eventually terminated to form defect pairs reminiscent of pattern
formation observed in other studies of active nematics
\cite{giomi11,FWZ2013} (Fig. \ref{numerics}c).

\begin{figure*}
 \begin{center}
  \includegraphics[width = 165mm]{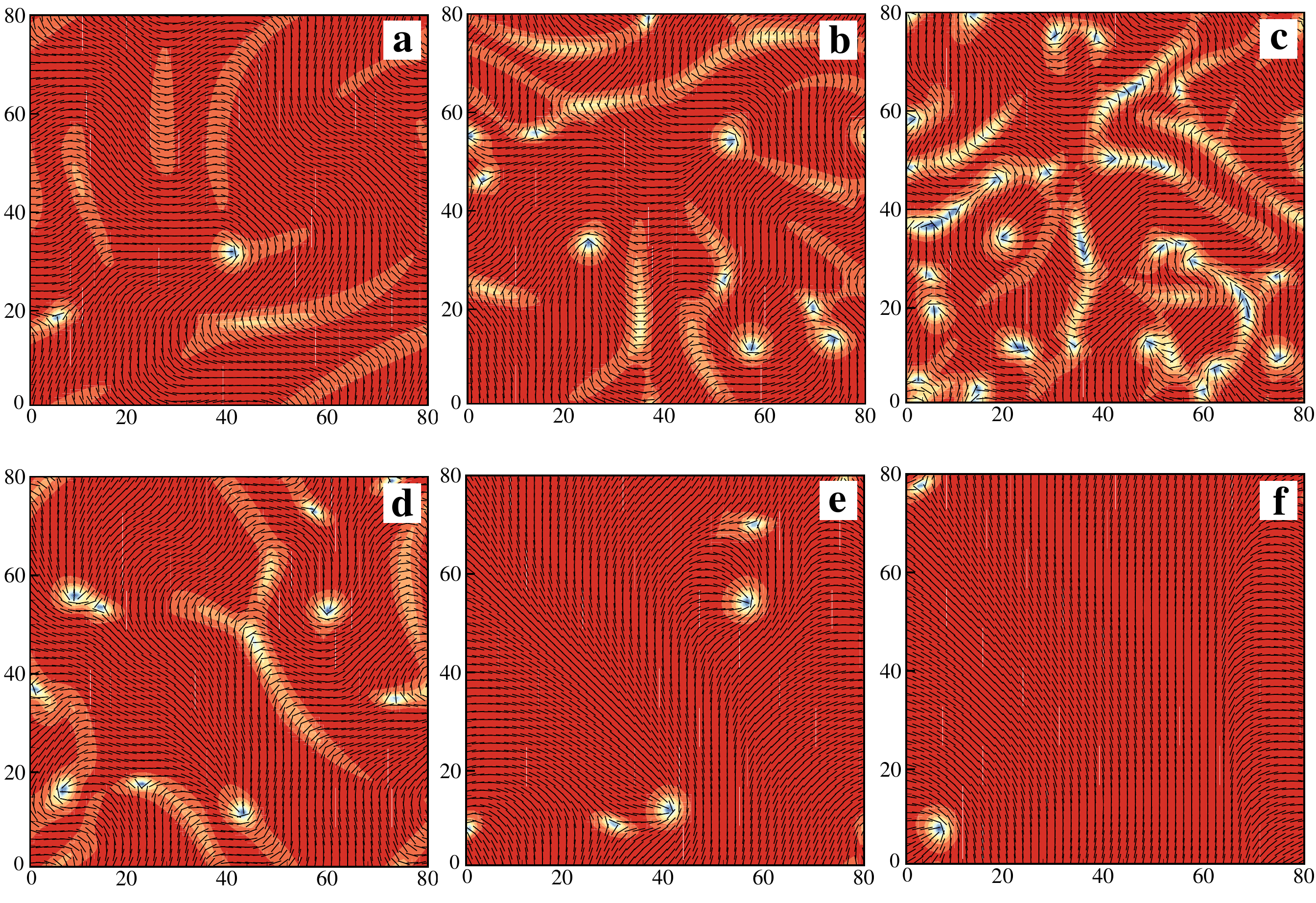}
 \end{center}
 \caption{Snapshots of motile defects when choosing different stress
   magnitudes.  For plots in (a-c), $\alpha_1 + \alpha_2 = -2.0$ is
   fixed: (a) $\alpha_1 = -2.0$ and $\alpha_2 = 0$; (b)
   $\alpha_1 = -1.0$ and $\alpha_2 = -1.0$; (c) $\alpha_1 = 0$ and
   $\alpha_2 = -2.0$.  (d-f) show time sequential snapshots
   ($t = 0, 100, 300$) of a case initially starting from the state
   with streaming defects but the fluid velocity field is turned off
   ($\bU = 0$). The system slowly relaxes to a globally
   nematic-aligned state. In (a-f), the color scale is the same as
   Fig. \ref{numerics}b.}
\label{alptest}
\end{figure*}

\subsubsection{Contributions of polar-specific active stresses}

Motivated by the linear stability analysis, we are able to tune the
system length scale by changing the magnitude of the two active
stresses. Generally, increasing either $\alpha_{aa}$ or $\alpha_{pa}$
increases the number of defect pairs, which shortens the
characteristic length scale in the dynamics
(Fig. \ref{instability}d-f). However, these two stresses arise from
different polar arrangements of MTs.  To understand the contributions
of the two active stresses, we fixed the total stress
$\alpha=\alpha_{aa}+\alpha_{pa}$ while varying the individual
$\alpha_{aa}$ and $\alpha_{pa}$ (Fig.~\ref{alptest}a-c). The case with
only polar-aligned active stress ($\alpha_{aa}=0$) produces more
defects than does the case with only anti-aligned active stress
($\alpha_{pa}=0$). Curiously, this seems due to the defects
themselves, as the passage of a $+1/2$-order defect leaves behind it
regions of higher polar order, and thus large bulk regions for
polar-aligned interactions. On the other hand, it appears that either
active stress ({\it aa} or {\it pa}) taken individually will produce
qualitatively similar flows and defect dynamics. Hence, the
qualitative nature of the large-scale dynamics does not by itself
isolate the precise origins of a destabilizing stress.

The linear stability analysis indicates that the instability is
coupled to the long-range hydrodynamic interactions. Consistent with
this, if we begin in a state with streaming defects and then turn off
the fluid flow by setting ${\bf U}=0$, defect creation stops while
defect annihilation continues, and the system relaxes to an aligned
nematic (Fig.~\ref{alptest}d-f).

\subsubsection{Formation of polar lanes}

\begin{figure*}
 \begin{center}
  \includegraphics[scale = 0.6]{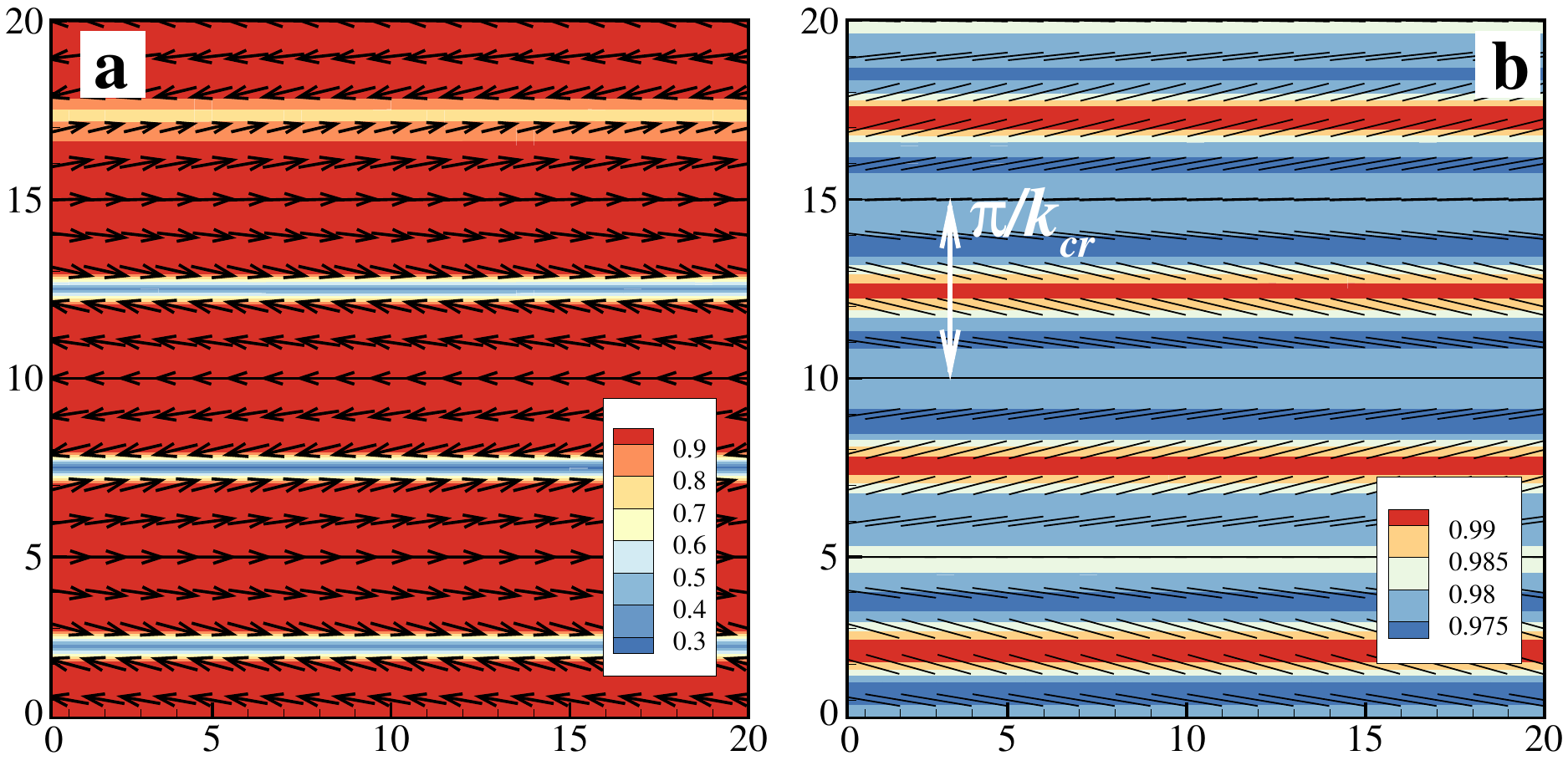}
 \end{center}
 \caption{Snapshots of polar lanes: (a) The polarity vector field
   $\bq$ superimposed upon its magnitude, and (b) the nematic director
   field $\bn$ superposed on the colormap of the scalar order
   parameter. }
\label{laning}
\end{figure*}

One apparent difference between the BD-kMC model and continuum kinetic
theory is the formation of polar lanes observed in the BD-kMC
model. To examine this discrepancy, we turned off hydrodynamics in our
continuum theory by forcing ${\bf U} = 0$. Then we find emergence of
polar lanes (Fig.~\ref{laning}) for simulations initialized in the
nematically-aligned state (Eq.~(\ref{alignedsol})) along the
$x-$direction.  The polarity field still aligns in $\hat{\bf x}$ but
forms oscillating switching bands in $\hat{\bf y}$, with relatively
small corresponding oscillations in the nematic director. Therefore,
removing hydrodynamic interactions leads to dynamics strikingly
different from the crack formation due to hydrodynamic instability
discussed above.  We performed numerical linear stability analysis (as
in Eqn.~\eqref{linearSmo}) for the laning instability.  This leads to
a characteristic length scale $\pi/k_{cr}$ that agrees well with the
lane size in nonlinear simulations (Fig.~\ref{laning}b).

These results suggest that the observation of polar lanes in the
BD-kMC model is related to the lack of long-range hydrodynamic
interactions in this model. We found that the laning instability has a
slow time scale compared to the hydrodynamic instabilities
(Fig.~\ref{instability}a); therefore, in the continuum kinetic theory
with long-range hydrodynamic interactions we observed the onset of
hydrodynamic instabilities rather than the laning instability.  This
laning instability is inherently polar, because it is driven by the
separation of the polarity field. Therefore, it arises from different
physics than similar patterns observed in apolar models
\cite{giomi11}.

\section{Conclusion}

We have developed a multiscale polar theory to describe a suspension
of MTs driven by the activity of plus-end directed motor protein
complexes (Fig.~\ref{schematic}).  First, we performed detailed BD-kMC
simulations that revealed polarity sorting and polar-specific active
stresses.  This model differs from previous simulation models of
motor-filament systems \cite{nedelec07,head11,head14} in the treatment
of motor binding and unbinding: previous work used simple binding
rules that do not obey the principle of detailed balance. Our more
accurate treatment of crosslink statistical mechanics is important to
determine how alterations from equilibrium motor distributions occur
due to nonequilibrium activity and the resulting active stresses
generated.

For MT densities that form an equilibrium 2D nematic in the absence of
motors, adding motor activity leads to MT flows driven by polarity
sorting and the formation of polar lanes (Figs.~\ref{figure02},
\ref{figure03}).  The mean-squared displacement of MT position becomes
superdiffusive and nearly ballistic at long times along the nematic
director. For polar-aligned pairs, the distribution of motor tether
extension shifts toward smaller extensions than in the equilibrium
case due to nonequilibrium tether relaxation; for anti-aligned pairs,
the distribution shifts toward positive extension due to oppositely
directed motor motion that drives polarity sorting.  MT displacement
distributions and instantaneous speeds along the nematic director are
strongly dependent on the local polar environment, consistent with the
continuum model: the instantaneous velocity depends approximately
linearly on the local polar environment, as expected when filament
movements are determined mainly by polarity sorting.

Over a wide range of parameters in the BD-kMC model, anisotropic
extensile stress is generated.  The stress produced per MT is largest
for filaments in anti-polar environments, suggesting that polarity
sorting is the dominant source of pairwise extensile stress. As
previously noted, if an MT pair begins sliding when the two minus-ends
touch and slides under a force proportional to pair overlap until the
two plus-ends meet, then the contractile motion would perfectly
balance the extensile motion and the total extensile stress would be
zero\cite{kruse00,Liverpool05,lenz12,lenz14}. Symmetry appears to be
broken in our BD-kMC simulations first, because MTs are unlikely to
begin interacting exactly when their minus ends are touching, and
second, because more motors are bound on average during extensional
motion (Fig.~\ref{figure03}).  Extensile stress is also generated for
polar-aligned pairs of MTs due to nonequilibrium motor tether
relaxation: the tether of a longitudinally stretched motor on parallel
filaments pulls back on the leading motor, slowing it, and pulls
forward on the trailing motor. Because the motor relaxes to become
more perpendicular to the filament pair, the motor-induced contractile
stress along the director is decreased, leading to a net anisotropic
extensile stress in the alignment direction. This is to our knowledge
a novel mechanism of extensile stress production\cite{GBGBS2015}. While
the per-pair extensile stress due to polar-aligned MT pairs is
typically smaller than for anti-aligned pairs, polarity sorting and
lane formation produce larger numbers of polar-aligned MT pairs than
of anti-aligned. In our BD-kMC simulations, which lack the effect of
hydrodynamics, the overall contributions of polar-aligned and
anti-aligned pairs to the extensile stress are comparable.

An interesting aspect of our BD-kMC study is that active stresses are
extensile, which is very different from the contractility observed in
actin-myosin gels\cite{bendix08}. Actin filament buckling appears to
be a key microscopic mechanism driving contractility in actin-myosin
systems\cite{silva11,lenz12,lenz12a,wang12,cordoba14,lenz14}.  The
greater rigidity of MTs in our model, the nematic ordering, and the
fluidity of MT motion may all contribute to extensile stress
generation in our system.

We incorporated the polar-specific active stresses into a kinetic
theory of Doi-Onsager type (Fig.~\ref{coarse}), and studied the effect
of hydrodynamic interactions. In the continuum model, we can derive
the origin and dependencies of extensile stresses driven by polarity
sorting. However, our understanding is less clear of the stress from
polar-aligned interactions, and a reduced continuum model of them
would be useful.  We find a streaming-nematic state similar to that
observed in recent experiments by Sanchez {\it et
  al.}\cite{sanchez12}. Defect pairs are born along incipient cracks
of low nematic order, and these cracks correlate with fluid jets
(Fig.~\ref{numerics}). The movement of $+1/2$ defects can be tracked
by associated vortex pairs, and strongly correlates with the active
force (Figs.~\ref{interface}, \ref{polarity}). We also identify a
hydrodynamic instability of nearly-aligned MTs that causes formation
of incipient cracks (see also Giomi {\it et al.}\cite{giomi11}) , and
hence serves as a source of complex dynamics (Figs.~\ref{psi},
\ref{instability}).

Our results are qualitatively similar to previous work studying liquid
crystal hydrodynamic driven by an apolar active stress
\cite{giomi13,giomi14,TGY2013,thampi14a,thampi14b,thampi14c}. These
works have improved our understanding of the speed of defect motion,
velocity correlations, and defect dynamics in active nematics, but
cannot address the microscopic origins of active stresses. Our work
seeks to connect these phenomena to MT/motor-protein interactions that
are intrinsically polar. Therefore, we link microscopic polar
interactions to macroscopic phenomena.  In future work it would be
interesting to further coarse-grain our kinetic model, say through a
moment closure approximation, and to generalize current apolar active
liquid crystal models to include polarity sorting and
polarity-dependent stresses.

Interestingly, in modeling the experiments of Sanchez {\it et al.}
\cite{sanchez12}, we find that by accounting for the outer fluid drag
on the immersed layer dynamics, we are able to determine a clear
characteristic length scale. This does not occur in active nematic
models based on bulk dynamics \cite{ESS2013,giomi13,TGY2013}, although
length selection has been reported for swimmers confined to immersed
thin films \cite{leoni10}, and substrate friction changes length-scale
selection in 2D active nematic models\cite{thampi14c}.

We find that either polar-specific active stress---associated with
polarity sorting or motor tether relaxation---taken individually
produces qualitatively similar flow and defect dynamics. Therefore
such dynamics alone do not isolate the origins of a destabilizing
stress.  Generally, increasing either $\alpha_{aa}$ (from anti-aligned
interactions) or $\alpha_{pa}$ (from polar-aligned interactions)
increases the number of defect pairs, which shortens the
characteristic length-scale in the dynamics
(Fig.~\ref{instability}). However, these two stresses arise from
different polar arrangements of MTs. Fixing
$\alpha=\alpha_{aa}+\alpha_{pa}$, the case with $\alpha_{aa}=0$ (only
polar-aligned active stress) produces more defects than does the case
with $\alpha_{pa}=0$ (only anti-aligned active stress). Curiously,
this seems due to the defects themselves, as the passage of a $+1/2$
defect leaves behind it regions of higher polar order, and thence
large bulk regions for polar-aligned interactions
(Fig.~\ref{alptest}). Perhaps the systematic experimental study of the
dynamics of photobleached regions would reveal which of these two
stresses, anti-aligned or polar-aligned, is actually dominant, or
whether the unstable dynamics arises from some other source.

One apparent difference of the kinetic theory with the BD-kMC model is
the polar laning evinced by the latter. To explore this we turned off
hydrodynamics in the kinetic model, since it is absent in the BD-kMC
model, and also found polar laning there (Fig.~\ref{laning}). Further,
as revealed by linear stability analysis of the kinetic model, there
is an instability to polar laning that occurs on a much longer time
scale (consistent with the time for lanes to emerge in the BD-kMC
model) than the hydrodynamic instabilities studied here. Additionally,
without long-range hydrodynamic interactions, the kinetic theory does
not show the persistent production of defects. It would be
illuminating to compare the kinetic theory with BD-kMC simulations
that incorporate hydrodynamic interactions. This might be done using
fast-summation methods applied earlier to the study of motile
suspensions \cite{SDS2005,saintillan12}.

{\it Acknowledgements.} We thank D. Chen and D. Needleman
for useful discussions. This work was funded by NSF grants
DMR-0820341 (NYU MRSEC: TG,MJS), DMS-0920930 (MJS), EF-ATB-1137822
(MB), DMR-0847685 (MB), and DMR-0820579 (CU MRSEC: MG); DOE grant
DE-FG02-88ER25053 (TG,MJS); NIH grant R01 GM104976-03 (MB,MJS); and
the use of the Janus supercomputer supported by NSF grant CNS-0821794.

\bibliography{zoterolibrary}
\bibliographystyle{unsrt}


\end{document}